\documentclass[journal,a4paper,10pt]{IEEEtran}

\usepackage[utf8]{inputenc} 
\usepackage[T1]{fontenc}
\usepackage{url}
\usepackage{ifthen}
\usepackage[noadjust]{cite}
\usepackage[cmex10]{amsmath}
\usepackage{multirow}
\usepackage{cite}
\usepackage{amsmath,amssymb,amsfonts}
\usepackage{algorithm}
\usepackage{algpseudocode}
\usepackage{graphicx}
\usepackage{makecell}
\usepackage{textcomp}
\usepackage{xcolor}
\usepackage{caption}
\usepackage{mathtools}
\usepackage{comment}
\usepackage{amsthm}
\usepackage{csquotes}
\usepackage{enumitem}
\usepackage{stfloats}
\usepackage{subfiles}
\newtheorem{thm}{Theorem}
\newtheorem{lem}{Lemma}

\newtheorem{rem}{Remark}

\newtheorem{defn}{Definition}

\DeclarePairedDelimiter\ceil{\lceil}{\rceil}
\DeclarePairedDelimiter\floor{\lfloor}{\rfloor}

\begin{document}
\title{Combinatorial Multi-Access Coded Caching: Improved Rate-Memory Trade-off with\\ Coded Placement}

\author{K. K. Krishnan Namboodiri, 
	\IEEEmembership{Graduate Student Member, IEEE} and B. Sundar Rajan, \IEEEmembership{Life Fellow, IEEE}
	\thanks{This work was supported partly by the Science and Engineering Research Board (SERB) of Department of Science and Technology (DST), Government of India, through J.C. Bose National Fellowship to Prof. B. Sundar Rajan, and by the Ministry of Human Resource Development (MHRD), Government of India, through Prime Minister’s Research Fellowship (PMRF) to K. K. Krishnan Namboodiri. Part of the content of this manuscript appeared in \textit{Proc. IEEE Inf. Theory Workshop (ITW)}, Apr. 2023, doi: 10.1109/ITW55543.2023.10161686 \cite{NaR5}.	
			
		K. K. Krishnan Namboodiri and B. Sundar Rajan are with the Department of Electrical Communication Engineering, IISc Bangalore, India (e-mail: krishnank@iisc.ac.in, bsrajan@iisc.ac.in).  }
}

\maketitle

\begin{abstract}
This work considers the combinatorial multi-access coded caching problem introduced in the recent work by Muralidhar \textit{et al.} [P. N. Muralidhar, D. Katyal, and B. S. Rajan, ``Maddah-Ali-Niesen scheme for multi-access coded caching,'' in \textit{IEEE Inf. Theory Workshop (ITW)}, 2021] The problem setting consists of a central server having a library of $N$ files and $C$ caches each with capacity $M$. Each user in the system can access a unique set of $r<C$ caches, and there exist users corresponding to every distinct set of $r$ caches. Therefore, the number of users in the system is $\binom{C}{r}$. For the aforementioned combinatorial multi-access setting, we propose a coded caching scheme with an MDS code-based coded placement. This novel placement technique helps to achieve a better rate in the delivery phase compared to the optimal scheme under uncoded placement when $M> N/C$. For a lower memory regime, we present another scheme with coded placement, which outperforms the optimal scheme under uncoded placement if the number of files is no more than the number of users. Further, we derive an information-theoretic lower bound on the optimal rate-memory trade-off of the combinatorial multi-access coded caching scheme. In addition, using the derived lower bound, we show that the first scheme is optimal in the higher memory regime, and the second scheme is optimal if $N\leq \binom{C}{r}$. Finally, we show that the performance of the first scheme is within a constant factor of the optimal performance, when $r=2$.
\end{abstract}
\IEEEpeerreviewmaketitle
\section{Introduction}
\label{intro}
With smartphones, mobile applications and expanded connectivity, people are consuming more video content than ever. This increase in mobile video consumption is taking a toll on the internet, which has seen data traffic increase many-fold in a few years. However, the high temporal variability of on-demand video services leaves the network underutilized during off-peak hours. Utilizing the channel resources by employing low-cost cache memory to store contents at the user end during off-peak times is an effective way to ease the network traffic congestion in peak hours. In the conventional caching scheme, the users' demands are met by filling the caches during the placement phase (in off-peak hours without knowing the user demands) and by transmitting the remaining requested file contents in the uncoded form during the delivery phase (in the peak hours after knowing the user demands). In coded caching, introduced in \cite{MaN}, it was shown that by employing coding in the delivery phase among the requested contents, it is possible to bring down the network load further compared to the conventional caching scheme. In \cite{MaN}, Maddah-Ali and Niesen considered a single server broadcast network, where the server is connected to $K$ users through an error-free shared link. The server has a library of $N$ files, and the user caches have a capacity of $M$ files, where $M\leq N$ (since each user is equipped with a dedicated cache memory, we refer to this network as the dedicated cache network). During the placement phase, the server stores file contents (equivalent to $M$ files) in the caches without knowing the user demands. In the delivery phase, each user requests a single file from the server. In the delivery phase, the server makes coded transmissions so that the users can decode their demanded files by making use of the cache contents and the coded transmissions. The goal of the coded caching problem is to jointly design the placement and the delivery phases such that the rate (the size of transmission) in the delivery phase is minimized. The scheme in \cite{MaN} achieved a rate $K(1-M/N)/(1+KM/N)$, which is later shown to be the optimal rate under uncoded placement if $N\geq K$ \cite{YMA,WTP}. In \cite{CFL}, it was shown that by employing coding among the contents stored in the placement phase in addition to the coding in the delivery phase, a better rate-memory trade-off could be achieved. By using coded placement, further improvement in the rate was obtained by the schemes in \cite{AmG,TiC,Vil1,ZhT,Vil2}. The works in \cite{STC,SeT,WLG,WBW,GhR}, came up with different converse bounds for the optimal rate-memory trade-off of the coded caching scheme for the dedicated cache network. 

Motivated by different practical scenarios, coded caching was extended to various network models, including multi-access networks \cite{HKD}, shared cache networks \cite{PUE} etc. In the multi-access coded caching (MACC) scheme proposed in \cite{HKD}, the number of users, $K$ was kept to be the same as the number of caches. Further, each user was allowed to access $r<K$ neighbouring caches in a cyclic wrap-around fashion. The coded caching scheme under the cyclic-wrap model was studied in \cite{ReK1,SPE,ReK2,MaR,SaR,CWLZC}. Those works proposed different achievable schemes, all restricted to uncoded cache placement. In \cite{NaR1}, the authors proposed a coded caching scheme with an MDS code-based coded placement technique and achieved a better rate-memory trade-off in the low memory regime compared to the schemes with uncoded placement. Further, by deriving an information-theoretic converse, the scheme in \cite{NaR1} was shown to be optimal in \cite{NaR2}. The MACC problem (in the cyclic wrap-around network) was studied by incorporating security and privacy constraints in \cite{NaR3,NaR4}. The first work that considered a multi-access network with more users than the caches is \cite{KMR}. The authors established a connection between the MACC problem and design theory, and obtained classes of MACC schemes from resolvable designs. The works in \cite{MKR1,DaR} also relied on design theory to obtain MACC schemes.

In this paper, we consider the combinatorial multi-access network model introduced in \cite{MKR2}, which consists of $C$ caches and $K=\binom{C}{r}$ users, where $r$ is the number of caches accessed by a user. The combinatorial MACC scheme presented in \cite{MKR2} achieves the rate $R=\binom{C}{t+r}/\binom{C}{t}$ at cache memory $M=Nt/C$, where $t\in\{1,2,\dots,C-r+1\}$. Here onwards, we refer to the scheme in \cite{MKR2} as the \textit{MKR scheme}. The optimality of the MKR scheme under uncoded placement was shown in \cite{BrE}. In addition to that, in \cite{BrE}, the system model in \cite{MKR2} was generalized to a setting where more than one user can access the same set of $r$ caches. In \cite{ChR}, the authors addressed the combinatorial MACC problem with privacy and security constraints.

Similar network topologies were considered in various cache aided settings, including two-hop networks \cite{JTLC,ZeY} and multi-server networks \cite{MGL}. 
The two-hop network considered in \cite{JTLC} and \cite{ZeY} consists of a server and $K$ users connected via a set of $h$ cache-aided relay nodes. Each user is connected to a distinct set of $\rho$ relay nodes, where $\rho<h$, and thus the number of user is $K=\binom{h}{\rho}$. Those users also have dedicated caches. The scheme in \cite{JTLC} followed an uncoded random placement strategy, whereas the scheme in \cite{ZeY} relied on MDS-coded placement. The main difference between the combinatorial MACC scheme and the coded caching scheme for the two-hop network is in the delivery phase. In the combinatorial MACC setting, the server communicates directly to the users over an error-free broadcast link. However, in two-hop networks, the server responds to the users’ requests by transmitting signals to each of the relay nodes. Then, each relay node utilizes its received signal and its cached contents to transmit unicast signals to each of its connected end users. In \cite{MGL}, a network model with $P$ servers and $K$ cache-aided users is considered. Each user can get connected to a random set of $\rho$ servers, where $\rho\leq P$. In order to ensure that any $\rho$  servers collectively store the entire file library, the file contents are stored in the servers after an MDS encoding.

\begin{figure}[t]
	\captionsetup{justification = centering}
	\captionsetup{font=small,labelfont=small}
	\begin{center}
		\captionsetup{justification = centering}
		\includegraphics[width = 0.95\columnwidth]{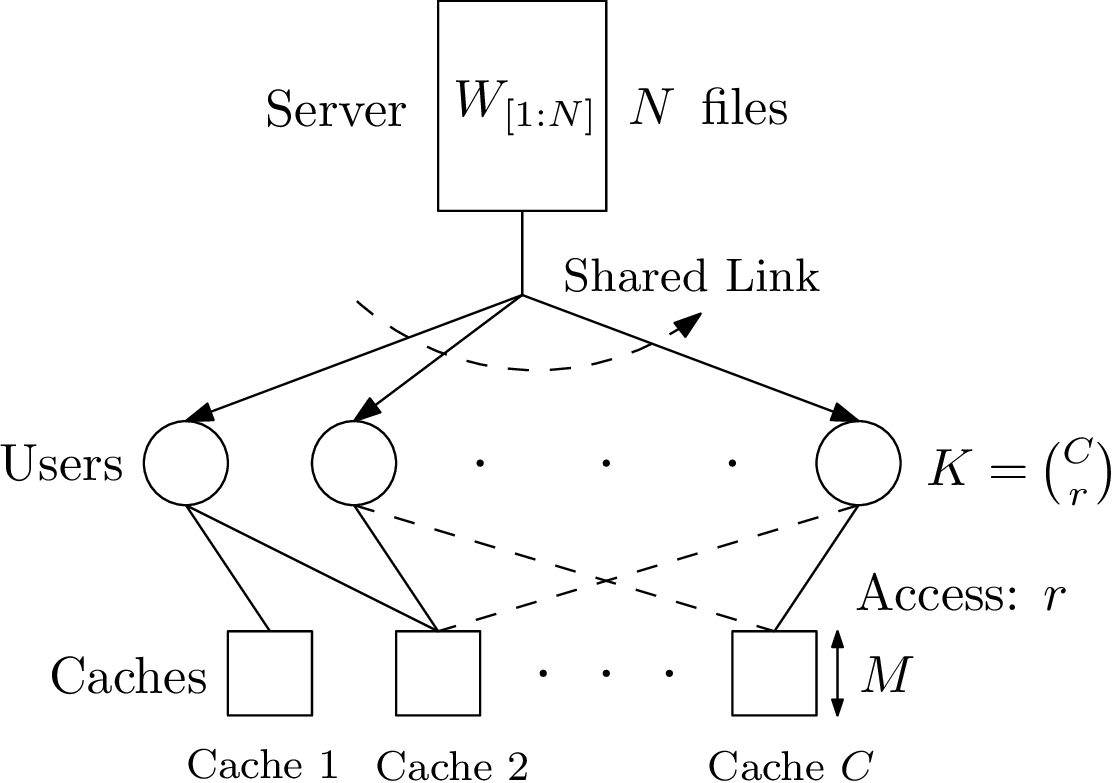}
		\caption{$(C,r,M,N)$ Combinatorial multi-access Network. }
		\label{network}
	\end{center}
\end{figure}

\subsection{Contributions}
In this paper, we study the combinatorial multi-access setting introduced in \cite{MKR2} and make the following technical contributions:
\begin{itemize}
	\item In \cite{MKR2}, the authors proposed a combinatorial MACC scheme (MKR scheme) that achieves the optimal rate under uncoded placement. However, in the MKR scheme, a user gets multiple copies of the same subfile from different caches that it accesses. In order to remove that redundancy, we introduce a novel, MDS code-based coded placement technique. By employing coded placement, we ensure that using the accessible cache contents, a user can decode all the subfiles that the corresponding user in the MKR scheme (by corresponding user, we mean that the user accessing the same set of caches) accesses in the uncoded form. The coded subfiles transmitted in the delivery phase of our proposed scheme (Scheme 1) remain the same as the transmissions in the MKR scheme. Thereby, our proposed scheme in Theorem \ref{thm:scheme1} achieves the same rate as the MKR scheme at a lesser cache memory value. For $M>N/C$, Scheme 1 outperforms the MKR scheme in terms of the rate achieved.

	\item For $0\leq M\leq N/C$, Scheme 1 has the same rate as the MKR scheme. Thus we present another achievability result for the combinatorial MACC scheme in Theorem \ref{thm:scheme2}. The new coding scheme (Scheme 2) achieves the rate $N-CM$ for $0\leq M\leq \frac{N-\binom{C-1}{r}}{C}$, which is strictly less than the rate achieved by the optimal scheme with uncoded placement when the number of files with the server is no more than the number of users in the system. In effect, by employing coding in the placement phase, we brought down the rate (except at $M=N/C$) further compared to the optimal scheme under uncoded placement.
	\item We derive an information-theoretic lower bound on the optimal rate-memory trade-off of the combinatorial MACC scheme (Theorem \ref{lowerbound}). To obtain this lower bound, we consider a scenario of decoding the entire server library at the user-end using multiple transmissions. In order to bound the uncertainty in those transmissions, we make use of an information-theoretic inequality (\textit{Han's inequality}) given in \cite{Han}. Han's inequality was originally used in the context of coded caching in \cite{STC,SeT}. Our bounding technique is similar to the approach used in \cite{STC} and \cite{NaR2}, where lower bounds were derived for the dedicated cache scheme and the MACC scheme with cyclic wrap-around, respectively. While deriving the lower bound, we do not impose any restriction on the placement phase. Thus, the lower bound is applicable even if the scheme employs coding in the placement phase.  
	
	\item By using the derived lower bound, we show that Scheme 1 is optimal for higher values of $M$, specifically when $\frac{M}{N}\geq \frac{\binom{C}{r}-1}{\binom{C}{r}}$ (Theorem \ref{optimality1}). Further, we show that Scheme 2 is optimal when $N\leq \binom{C}{r}$ (Theorem \ref{optimality2}). In addition, we show that when $r=2$ and $N\geq K$, the rate-memory trade-off given by Scheme 1 is within a constant multiplicative factor, 21, of the information-theoretic optimal rate-memory trade-off (Theorem \ref{r=2gap}). 
	
\end{itemize}


\subsection{Notations}
For a positive integer $n$, $[n]$ denotes the set $ \left\{1,2,\hdots,n\right\}$. For two positive integers $a,b$ such that $a\leq b$, $[a:b] = \{a,a+1,\hdots,b\}$. For two non-negative integers $n,m$, we have $\binom{n}{m} =\frac{n!}{m!(n-m)!}$, if $n\geq m$, and $\binom{n}{m}=0$ if $n<m$. Uppercase letters (excluding the letters $C,K,M,N$ and $R$) are used to denote random variables (the letters $C,K,M,N$ and $R$ are reserved for denoting system parameters). The set of random variables $\{V_a,V_{a+1},\dots,V_b\}$ is denoted as $V_{[a:b]}$. Calligraphic letters are used to denote sets. Further, for a set of positive integers $\mathcal{I}$, $V_{\mathcal{I}}$ represents a set of random variables indexed by the elements in $\mathcal{I}$ (for instance, $V_{\{2,4,5\}}=\{V_2,V_4,V_5\}$). Bold lowercase letters represent vectors, and bold uppercase letters represent matrices. The identity matrix of size $n\times n$ is denoted as $\mathbf{I}_n$. Further, $\mathbb{F}_q$ represents the finite field of size $q$. Finally, for a real number $z$, $(z)^+=\max(0,z)$.

The rest of the paper is organized as follows. Section \ref{system} describes the system model, and Section \ref{prelims} describes useful preliminaries. The main results on the achievability and converse are presented in Section \ref{main}. The proofs of the main results are given in Section \ref{proofs}.
\section{System Model and Problem Formulation}
\label{system}
The system model as shown in Fig. \ref{network} consists of a central server with a library of $N$ independent files, $W_{[1:N]}\triangleq \{W_1,W_2,\dots,W_N\}$, each of size 1 unit (we assume that 1 unit of a file is constituted by $f$ symbols from the finite field of size $q$)\footnote{We assume that $q$ is such that all MDS codes considered in this work exist over $\mathbb{F}_q$.}. There are $C$ caches, each with capacity $M$ units ($0\leq M\leq N$). i.e., each cache can store contents equivalent to $M$ files. There are $K$ users in the system, and each of them can access $r$ out of the $C$ caches, where $r<C$. We assume that there exists a user corresponding to any choice of $r$ caches from the total $C$ caches. Thus the number of users in the system is $K=\binom{C}{r}$. Further, each user is denoted with a set $\mathcal{U}$, where $\mathcal{U}\subset [C]$ such that $|\mathcal{U}|=r$. That is, a user is indexed with the set of caches that it accesses. In other words, user $\mathcal{U}$ has access to all the caches in the set $\mathcal{U}$. A system under the aforementioned setting is called the $(C,r,M,N)$ combinatorial multi-access network. The coded caching scheme under this model was introduced in \cite{MKR2}.

The $(C,r,M,N)$ combinatorial MACC scheme works in two phases, the placement phase and the delivery phase. In the placement phase, the server populates the caches with the file contents. The cache placement is done without knowing the demands of the users. The placement can be either coded or uncoded. By uncoded placement, we mean that files are split into subfiles and kept in the caches as such, while coded placement means that coded combinations of the subfiles are allowed to be kept in the caches. The number of subfiles into which a file is split is termed the subpacketization number. The contents stored in cache $c$, $c\in [C]$, is denoted as $Z_c$. In the delivery phase, user $\mathcal{U}$ requests file $W_{d_\mathcal{U}}$ from the server, where $d_\mathcal{U}\in [N]$, and the demand vector $\mathbf{d} = (d_\mathcal{U}:\mathcal{U}\subset [C],|\mathcal{U}|=r)$. In the demand vector, we arrange the users (subsets of size $r$) in the lexicographic order. Corresponding to the demand vector, the server makes a transmission $X$ of size $R$ units. We assume that the broadcast channel from the server to the users is error-free. The non-negative real number $R$ is said to be the rate of transmission. Finally, user $\mathcal{U}$ should be able to decode $W_{d_\mathcal{U}}$ using transmission $X$ and the accessible cache contents $Z_c,c\in \mathcal{U}$. That is, for $\mathcal{U}\subset [C]$ such that $|\mathcal{U}|=r$, we have 
\begin{equation}
\label{decodability}
H(W_{d_\mathcal{U}}|Z_\mathcal{U},X)=0
\end{equation}
where $Z_\mathcal{U} =\{Z_c: c\in \mathcal{U}\}$.
\begin{defn}
	For the $(C,r,M,N)$ combinatorial MACC scheme, a rate $R$ is said to be achievable if the scheme satisfies \eqref{decodability} with a rate less than or equal to $R$ for every possible demand vector. Further, the optimal rate-memory trade-off is
	\begin{equation}
	R^*(M) =\inf \{R: R \text{ is achievable}\}.
	\end{equation}
\end{defn}
The coded caching problem aims to design the placement and the delivery phases jointly so that the rate is minimized. In this work, we present two achievability results and a lower bound on $R^*(M)$.
\section{Preliminaries}
\label{prelims}
In this section, we discuss the preliminaries required for describing the coding scheme as well as the converse bound.
\subsection{Review of the combinatorial MACC scheme in \cite{MKR2} (MKR scheme)}
In the sequel, we describe the combinatorial MACC scheme presented in \cite{MKR2}. 

In the placement phase, the server divides each file into $\binom{C}{t}$ non-overlapping subfiles of equal size, where $t\triangleq CM/N\in [C]$. The subfiles are indexed with $t$-sized subsets of $[C]$. Therefore, we have, $W_n = \{W_{n,\mathcal{T}}:\mathcal{T}\subseteq [C],|\mathcal{T}|=t\}$ for all $n\in [N]$. The server fills cache $c$ as follows:
\begin{equation}
	Z_c = \left\{W_{n,\mathcal{T}}:\mathcal{T}\ni c, \mathcal{T}\subseteq [C],|\mathcal{T}|=t,n\in [N]\right \}
\end{equation} 
for every $c\in [C]$. According to the above placement, the server stores $\binom{C-1}{t-1}$ subfiles of all the files in a cache. Thus, we have, $M/N =\binom{C-1}{t-1}/\binom{C}{t}=t/C$. Now, assume that user $\mathcal{U}$ demands file $W_{d_\mathcal{U}}$ from the server. In the delivery phase, the server makes coded transmissions corresponding to every $\mathcal{S}\subseteq [C]$ such that $|\mathcal{S}|=t+r$. The server transmission is as follows:
\begin{equation}
	\bigoplus_{\substack{\mathcal{U}\subseteq \mathcal{S}\\|\mathcal{U}|=r}} W_{d_\mathcal{U},\mathcal{S}\backslash \mathcal{U}}.
\end{equation}
Therefore, the number of coded subfiles transmitted is $\binom{C}{t+r}$, where each coded subfile is $1/\binom{C}{t}$ of a file size. Thus the rate of transmission is $\binom{C}{t+r}/\binom{C}{t}$. 

Notice that user $\mathcal{U}$ gets subfile $W_{n,\mathcal{T}}$ for all $n\in [N]$ from the cache placement if $\mathcal{U}\cap \mathcal{T}\neq \phi$. Now, let us see how does the user get subfile $W_{d_\mathcal{U},\mathcal{T}}$ if $\mathcal{U}\cap \mathcal{T}= \phi$. Consider the transmission corresponding to the $(t+r)$-sized set $\mathcal{S}=\mathcal{U}\cup \mathcal{T}$. In the coded message
\begin{equation*}
	\bigoplus_{\substack{\mathcal{U}\subseteq \mathcal{S}\\|\mathcal{U}|=r}} W_{d_\mathcal{U},\mathcal{S}\backslash \mathcal{U}}=W_{d_\mathcal{U},\mathcal{T}}\oplus \bigoplus_{\substack{\mathcal{U'}\subseteq \mathcal{S}\\|\mathcal{U}'|=r\\\mathcal{U}'\neq \mathcal{U}}} W_{d_{\mathcal{U}'},\mathcal{S}\backslash \mathcal{U}'}
\end{equation*}
user $\mathcal{U}$ has access to $W_{d_{\mathcal{U}'},\mathcal{S}\backslash \mathcal{U}'}$ for every $\mathcal{U}'\neq \mathcal{U}$, since $|\mathcal{U}\cap\mathcal{S}\backslash \mathcal{U}'|\neq 0$.  Therefore, user $\mathcal{U}$ can decode the demanded file $W_{d_\mathcal{U}}$.

\subsection{Maximum distance separable (MDS) codes \cite{MaS}}
An $[n,k]$ maximum distance separable (MDS) code is an erasure code that allows recovering the $k$ message/information symbols from any $k$ out of the $n$ coded symbols. Consider a systematic $[n,k]$ MDS code (over the finite field $\mathbb{F}_q$) generator matrix $\mathbf{G}_{k\times n} = [\mathbf{I}_k|\mathbf{P}_{k\times n-k}]$. Then, we have
\begin{align*}
[m_1,m_2,\dots,m_k,c_1,c_2,\dots,c_{n-k}]=[m_1,m_2,\dots,m_k]\mathbf{G}
\end{align*}
where the message vector $[m_1,m_2,\dots,m_k]\in \mathbb{F}_q^k$.

\subsection{Han's Inequality \cite{Han}}
To derive a lower bound on $R^*(M)$, we use the following lemma which gives an inequality on subset of random variables.
\begin{lem}[Han's Inequality \cite{Han}]
	\label{Entropy}
	Let $V_{[1:m]}=\{V_1,V_2,\dots,V_m\}$ be a set of $m$ random variables. Further, let $\mathcal{A}$ and $\mathcal{B}$ denote subsets of $[1:m]$ such that $|\mathcal{A}|=a$ and $|\mathcal{B}|=b$ with $a\geq b$. Then, we have 
	\begin{equation}
	\frac{1}{\binom{m}{a}}\sum_{\substack{\mathcal{A}\subseteq [1:m] \\ |\mathcal{A}|=a}} \frac{H(V_{\mathcal{A}})}{a}\leq \frac{1}{\binom{m}{b}}\sum_{\substack{\mathcal{B}\subseteq [1:m] \\ |\mathcal{B}|=b}} \frac{H(V_{\mathcal{B}})}{b}
	\end{equation}
	where $V_{\mathcal{A}}$ and $V_{\mathcal{B}}$ denote the set of random variables indexed by the elements in $\mathcal{A}$ and $\mathcal{B}$, respectively.
\end{lem}
\subsection{Motivating Example}
\label{motex}
Even though the MKR scheme is optimal under uncoded placement, with an example, we show that further reduction in rate is possible with the help of coded placement. 
\begin{figure}[t]
	\captionsetup{justification = centering}
	\captionsetup{font=small,labelfont=small}
	\begin{center}
		\captionsetup{justification = centering}
		\includegraphics[width = 0.95\columnwidth]{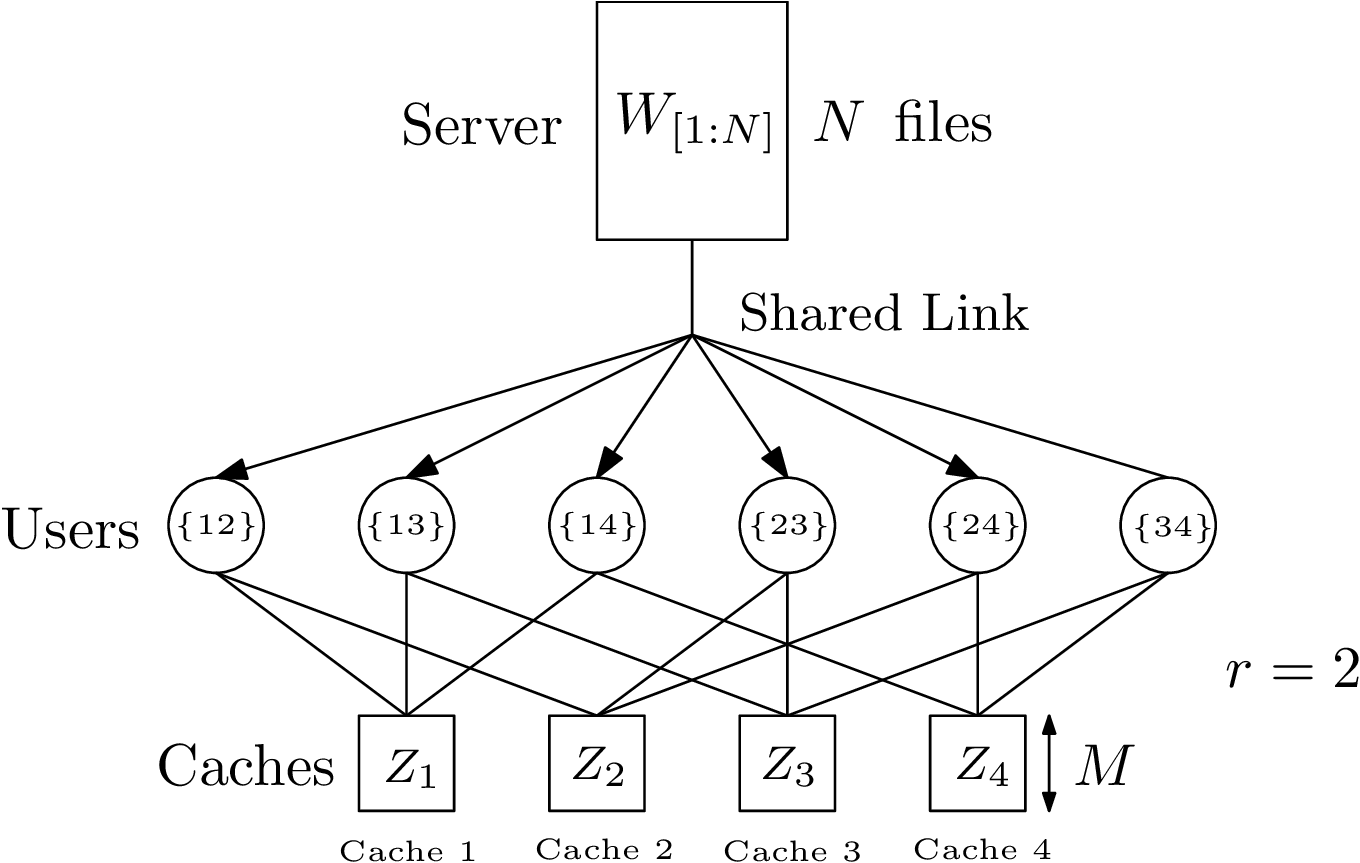}
		\caption{$(4,2,M,N)$ Combinatorial multi-access Network. }
		\label{examplefigure}
	\end{center}
\end{figure}

Consider the $(C=4,r=2,M=3,N=6)$ combinatorial multi-access network in Fig. \ref{examplefigure}. There are $\binom{4}{2}=6$ users in the system, each denoted with a set $\mathcal{U}$, where $\mathcal{U} \subset [C]$ such that $|\mathcal{U}|=2$. Let us first see the MKR scheme, where the placement is uncoded. In the placement phase, the server divides each file into 6 non-overlapping subfiles of equal size. Further, each subfile is denoted with a set $\mathcal{T}$, where $\mathcal{T}\subset[C]$ such that $|\mathcal{T}|=2$. Thus, we have $W_n =\{W_{n,12},W_{n,13},W_{n,14},W_{n,23},W_{n,24},W_{n,34}\}$ for all $n\in [6]$. Note that, even though the subfile indices are sets, we omitted the curly braces for simplicity. The server populates the $k^{\text{th}}$ cache as follows:
\begin{equation*}
Z_k = \{W_{n,\mathcal{T}}: k\in \mathcal{T},\mathcal{T}\subset[C], |\mathcal{T}|=2\}.
\end{equation*}
The contents stored in the caches are given in Table \ref{exampleplacement}. 
From the cache placement, user $\mathcal{U}$ has access to all the subfiles (of every file) except those indexed with 2-sized subsets of $[C]\backslash \mathcal{U}$. For example, user $\{1,2\}$ has access to all the subfiles except subfile $W_{n,34}$.

\begin{table*}[h!]
\begin{center}
		\resizebox{0.55\textwidth}{!}{
			\begin{tabular}{|c|c|c|c|}
				\hline 
				Cache 1 &Cache 2 &Cache 3 &Cache 4 \\
				\hline
				\hline
				$W_{n,12}$ &
				 $W_{n,12}$ &
				 $W_{n,13}$ & 
				 $W_{n,14}$\\				
				$W_{n,13}$ &
				 $W_{n,23}$ & 
				 $W_{n,23}$ &
				  $W_{n,24}$\\
				$W_{n,14}$ &
				 $W_{n,24}$ & 
				 $W_{n,34}$ &
				  $W_{n,34}$\\
				\hline
			\end{tabular}
		}
		\caption{$(C=4,r=2,M=3,N=6)$:  MKR scheme placement (subfiles are stored $\forall n\in [6]$)}
		\label{exampleplacement}
	\end{center}
\end{table*}

\begin{table*}[ht!]
	\begin{center}
		\resizebox{0.8\linewidth}{!}{
			\begin{tabular}{|c|c|c|c|}
				\hline 
				Cache 1 &Cache 2 &Cache 3 &Cache 4 \\
				\hline
				\hline
				$W_{n,12}^{(1)}$ & 
				$W_{n,12}^{(2)}$ & 
				$W_{n,13}^{(2)}$ &
				$W_{n,14}^{(2)}$\\
				$W_{n,13}^{(1)}$ &
				 $W_{n,23}^{(1)}$ & 
				 $W_{n,23}^{(2)}$ & 
				 $W_{n,24}^{(2)}$\\
				$W_{n,14}^{(1)}$ &
				 $W_{n,24}^{(1)}$ & 
				 $W_{n,34}^{(1)}$ &
				  $W_{n,34}^{(2)}$\\
				$W_{n,12}^{(2)}+W_{n,13}^{(2)}$ &
				 $W_{n,12}^{(1)}+W_{n,23}^{(2)}$ &
				  $W_{n,13}^{(1)}+W_{n,23}^{(1)}$ &
				   $W_{n,14}^{(1)}+W_{n,24}^{(1)}$\\
				$W_{n,13}^{(2)}+W_{n,14}^{(2)}$ &
				 $W_{n,23}^{(2)}+W_{n,24}^{(2)}$ &
				  $W_{n,23}^{(1)}+W_{n,34}^{(2)}$ & 
				  $W_{n,24}^{(1)}+W_{n,34}^{(1)}$\\
				\hline
			\end{tabular}
		}
		\caption{$(C=4,r=2,M=2.5,N=6)$: Cache contents (subfiles are stored $\forall n\in [6]$)}
		\label{examplecodedplacement}
	\end{center}
\end{table*}

In the delivery phase, the users reveal their demands. Let $W_{d_\mathcal{U}}$ be the file demanded by user $\mathcal{U}$. Then the server transmits $W_{d_{12},34}+W_{d_{13},24}+W_{d_{14},23}+W_{d_{23},14}+W_{d_{24},13}+W_{d_{34},12}$. The claim is that all the users will get their demanded file from the above placement and delivery phases. For example, user $\{1,2\}$ has access to all the subfiles except subfile $W_{n,34}$. However, user $\{1,2\}$ can get $W_{d_{12},34}$ by subtracting the remaining subfiles from the transmitted message. Similarly, it can be verified that all the users will get their demanded files. Thus the rate achieved is 1/6. This achieved rate is, in fact, optimal under uncoded placement \cite{BrE}. Notice that every subfile is cached in two caches. For instance, subfile $W_{n,12}$ is cached in cache 1 and cache 2. Thus user $\{1,2\}$ who accesses those two caches get multiple copies of the same subfile. This leads to the question of whether we can do better than the MKR scheme. 

Now, we employ a coded placement technique and achieve the same rate 1/6 at $M=2.5$. The coding scheme is described hereinafter. The server divides each file into 6 subfiles of equal size, and each subfile is indexed with a set $\mathcal{T}$, where $\mathcal{T} \subset [C]$ such that $|\mathcal{T}|=2$. Thus, we have  $W_n =\{W_{n,12},W_{n,13},W_{n,14},W_{n,23},W_{n,24},W_{n,34}\}$ for all $n\in [6]$. Further, each subfile is divided into 2 mini-subfiles both having half of a subfile size, we have $W_{n,\mathcal{T}} =\{W_{n,\mathcal{T}}^{(1)},W_{n,\mathcal{T}}^{(2)}\}$ for all subfiles. The cache placement is summarized in Table \ref{examplecodedplacement}. Each cache contains 5 mini-subfiles (3 uncoded and 2 coded mini-subfiles), each of size $1/12$ of a file size. Thus, the size of the caches is $2.5$.

Let $W_{d_\mathcal{U}}$ be the file demanded by user $\mathcal{U}$ in the delivery phase. Then the server transmits $W_{d_{12},34}+W_{d_{13},24}+W_{d_{14},23}+W_{d_{23},14}+W_{d_{24},13}+W_{d_{34},12}$. Notice that, this coded message is the same as the transmitted message in the MKR scheme. Thus the rate achieved is also the same as the MKR scheme, which is 1/6. Now, it remains to show that the users can decode their demanded files. Since our scheme follows the same delivery policy as the MKR scheme, it is enough to show that a user in our scheme can decode the subfiles, which are accessible for the corresponding user (user accessing the same set of caches) in the MKR scheme. Consider user $\mathcal{U}$ who accesses the cache $k$ for all $k\in \mathcal{U}$. Consider user $\{1,2\}$ accessing cache 1 and cache 2. The user gets $W_{n,12}^{(1)}$ from cache 1 and $W_{n,12}^{(2)}$ from cache 2, and thus it obtains the subfile $W_{n,12}=\{W_{n,12}^{(1)},W_{n,12}^{(2)}\}$ for all $n\in [6]$. Using $W_{n,12}^{(2)}$, the user can decode $W_{n,13}^{(2)}$ and $W_{n,14}^{(2)}$ from cache 1. Similarly, from cache 2, the subfiles $W_{n,23}^{(2)}$ and $W_{n,24}^{(2)}$ can be decoded using $W_{n,12}^{(1)}$. That means, user $\{1,2\}$ obtains the subfiles $W_{n,12},W_{n,13},W_{n,14},W_{n,23}$ and $W_{n,24}$ for every $n\in [6]$. From Table \ref{examplecodedplacement}, it can be verified that a user $\mathcal{U}$ gets all the subfile $W_{n,\mathcal{T}}$ such that $\mathcal{T}\cap \mathcal{U}\neq \phi$. Thus a user can decode the subfiles, which are accessible for the corresponding user in the MKR scheme in the placement phase. In essence, by employing a coded placement technique, we could achieve the same rate, but for a smaller cache memory value.  

Next, we show that for the $(C=4,r=2,M=3,N=6)$ coded caching scheme, it is possible to meet all the user demands without the server transmission. In other words, for the $(C=4,r=2,M=3,N=6)$ coded caching scheme, rate $R=0$ is achievable. In the placement phase, each file is divided into 2 subfiles of equal size, $W_n=\{W_{n,1},W_{n,2}\}$ for all $n\in [6]$. Then encode $(W_{n,1},W_{n,2})$ with a $[4,2]$ MDS code generator matrix $\mathbf{G}_{2\times 4}$. Thus, we have the coded subfiles  
\begin{equation*}
(C_{n,1},C_{n,2},C_{n,3},C_{n,4}) = (W_{n,1},W_{n,2})\mathbf{G}_{2\times 4}.
\end{equation*}    
Then the server fills the caches as follows: $Z_1 = \{C_{n,1};n\in[6]\}$, $Z_2 = \{C_{n,2};n\in[6]\}$, $Z_3 = \{C_{n,3};n\in[6]\}$, and $Z_4 = \{C_{n,4};n\in[6]\}$. From this placement, one user gets two coded subfiles from the caches (one from each cache). Since, any two columns of matrix $\mathbf{G}$ are independent, users can decode all the files, without further transmissions.
\section{Main Results}
\label{main}
In this section, we present two achievability results for the $(C,r,M,N)$ combinatorial MACC scheme. Further, we derive an information-theoretic lower bound on the optimal rate-memory trade-off, $R^*(M)$.

\begin{thm}
	\label{thm:scheme1}
	Let $t\in [C-r+1]$. Then, for the $(C,r,M,N)$ combinatorial MACC scheme, the rate $\binom{C}{t+r}/\binom{C}{t}$ is achievable at cache memory
	\begin{equation}
	\label{thm1eqn}
	M=
	N\left(\frac{t}{C}-\frac{1}{\binom{C}{t}}\sum\limits_{i=1}^{\tilde{r}-1}\frac{\tilde{r}-i}{r}\binom{r}{\tilde{r}-i+1}\binom{C-r}{t-\tilde{r}+i-1}\right)\\
	\end{equation}
	where $\tilde{r}=\min(r,t)$.
\end{thm}
Proof of Theorem \ref{thm:scheme1} is given in Section \ref{proof:thm1}. \hfill$\blacksquare$

In Section \ref{proof:thm1}, for the $(C,r,M,N)$ combinatorial MACC problem, we explicitly present a coding scheme making use of coded placement. As we have seen in the example in Section \ref{motex}, coding in the placement phase helps to avoid redundancy in the cached contents (accessed by a user) in the MKR scheme. The example shows that by storing fewer coded subfiles in the placement phase, a user can decode the same subfiles as the corresponding user in the MKR scheme (by \textit{corresponding user}, we mean that the user accessing the same set of caches) gets in the uncoded form. In other words, it is possible to achieve the same rate as the MKR scheme by using a lesser value of cache memory by employing coding in the placement. For a cache memory value $M$ as defined in \eqref{thm1eqn} corresponds to a $t\in [C-r]$, we can modify the rate expression as (proof is given in Appendix \ref{AppendixC})
\begin{equation}
\label{rate}R(M) = \frac{\binom{C}{t+r}}{\binom{C}{t}} = \frac{\binom{C}{r}(1-\frac{rM}{N})}{\binom{t+r}{r}}.
\end{equation} 
Also, the parameter $t=C-r+1$ corresponds to $M=N/r$ (proof is provided in Appendix \ref{AppendixA}) and rate, $R(M=N/r)=0$. Therefore, the rate expression \eqref{rate} is valid for every $t\in [C-r+1]$. For a general $0\leq M\leq N/r$, a lower convex envelope of these points is achievable via memory sharing technique. In the rate expression, the term $(1-rM/N)$ shows that a user can access or decode $rM/N$ fraction of every file from the caches that it accesses and requires only the remaining portion of the required file from the delivery phase. Further, the denominator $\binom{t+r}{r}$ is the number of users simultaneously served by a coded transmission in the delivery phase. 

The parameter $t$ in the MKR scheme represents the number of times the entire server library is duplicated among the caches. 
In the MKR scheme, user $\mathcal{U}$ gets $|\mathcal{T}\cap \mathcal{U}|$ copies of a subfile $W_{n,\mathcal{T}}$. Therefore, a user gets a different number of copies of different subfiles depending upon the cardinality of the intersection between the user index set and the subfile index set. Our objective is to design a coded placement phase such that from the accessible cache contents, a user should be able to decode the subfiles that the corresponding user in the MKR scheme gets in the uncoded form. The main challenge of designing such a placement phase is that a user should be able to decode all the subfiles with a non-zero intersection between the user index set and the subfile index set. To tackle this challenge, we employ a two-level MDS encoding. First, we encode the subfiles with an MDS code generator matrix. Then, in each round of the placement phase, a few selected coded subfiles are encoded again with another MDS code generator matrix. This two-level encoding technique is different from the MDS code-based coded placement strategies followed in the schemes in \cite{ZeY,MGL} for two-hop networks and multi-server networks, respectively. In both cases, the entire file library was encoded with an MDS code. In our case, user $\mathcal{U}$, $\mathcal{U}\subseteq [C],|\mathcal{U}|=r$ should be able to decode all the subfiles $W_{n,\mathcal{T}}$ such that $\mathcal{T}\cap \mathcal{U}\neq \phi$, from the accessible cache contents. To ensure that, in general, a single-level encoding is not sufficient. Our first-level MDS encoding duplicates the file library, whereas our second-level MDS encoding ensures that the users can decode the required subfiles from the cache contents.

Further, note that the parameter $\tilde{r}=\min(r,t)$ is the maximum value of $|\mathcal{T}\cap \mathcal{U}|$ for any $\mathcal{T}$ and $\mathcal{U}$. In other words, $\tilde{r}$ is the maximum number of copies of a single subfile that is accessed by a user in the MKR scheme. Therefore, our placement phase depends upon $\tilde{r}$. In fact, the placement phase consists of $\tilde{r}$ rounds. i.e, Round 0, Round 1, $\dots$, Round $\tilde{r}-1$. The transmissions in our delivery phase is the same as that of the MKR scheme. Thus the coded placement phase should enable user $\mathcal{U}$ to decode all the subfiles $W_{n,\mathcal{T}}$ with $\mathcal{T}\cap \mathcal{U}\neq \phi$ from the content stored in the accessible caches. Let us denote the content stored in cache $c$ in Round $b$ as $Z_c^b$, where $c\in [C]$ and $b\in \{0\}\cup[\tilde{r}-1]$. Then, we design our placement phase such that user $\mathcal{U}$ be able to decode a subfile $W_{n,\mathcal{T}}$ with $|\mathcal{T}\cap\mathcal{U}|=\tilde{r}-\beta$ using $Z_c^0,Z_c^1,\dots,Z_c^\beta$ for every $c\in \mathcal{U}$. For instance, using $Z_c^0$ alone (for every $c\in \mathcal{U}$), the user can decode $W_{n,\mathcal{T}}$ if $|\mathcal{T}\cap\mathcal{U}|=\tilde{r}$. In further decoding, the user also uses the already decoded subfiles. Thus the cache content decoding happens sequentially. First, the user decodes the subfiles $W_{n,\mathcal{T}}$ with $|\mathcal{T}\cap\mathcal{U}|=\tilde{r}$, then the subfiles $W_{n,\mathcal{T}}$ with $|\mathcal{T}\cap\mathcal{U}|=\tilde{r}-1$ and so on. Finally, the user can decode the subfiles $W_{n,\mathcal{T}}$ with $|\mathcal{T}\cap\mathcal{U}|=1$.

 When $t=1$, there is no duplication in cached contents, and thus the contents accessed by a user from different caches are distinct. Therefore, performance improvement by coding is available when $t>1$. Further, the placement phase of the MKR scheme is independent of $r$, whereas our scheme takes $r$ into consideration during the cache placement. Also, our scheme needs to divide each file into $\tilde{r}!\binom{C}{t}$, whereas the MKR scheme needs a lesser subpacketization of $\binom{C}{t}$.

The MKR scheme achieves the rate $\binom{C}{t+r}/\binom{C}{t}$ at $M=Nt/C$, which is optimal under uncoded placement. Since $\sum\limits_{i=1}^{\tilde{r}-1}\frac{\tilde{r}-i}{r}\binom{r}{\tilde{r}-i+1}\binom{C-r}{t-\tilde{r}+i-1}$ is always positive for $t>1$, we achieve the same rate $\binom{C}{t+r}/\binom{C}{t}$ at $M<Nt/C$. This advantage is enabled with the help of coding in the placement phase. If $t=1$, we have cache memory $M=N/C$. For $0\leq M\leq N/C$, the scheme in Theorem \ref{thm:scheme1} has the same rate-memory trade-off as the MKR scheme. For every $M>N/C$, our proposed scheme performs strictly better than the MKR scheme in terms of the rate achieved.
\begin{rem}
	To quantify the improvement brought in by Theorem \ref{thm:scheme1} compared to the MKR scheme, we consider two memory points: a) corresponding to $t=2$ (a lower cache memory value), and b) corresponding to $t=C-r$ (a higher cache memory value). \\
	a) The parameter $t=2$ corresponds to $M = \frac{N}{C}(2-\frac{r-1}{C-1})$. The corresponding rate from Theorem \ref{thm:scheme1} is $R(M) = \frac{\binom{C}{r+2}}{\binom{C}{2}}$. But, the MKR scheme achieves the rate $R_{\text{uncoded}}^*(M)=(1+\frac{r(C+1)(r-1)}{2(C-r-1)(C-1)})\frac{\binom{C}{r+2}}{\binom{C}{2}}$. Therefore, at $M = \frac{N}{C}(2-\frac{r-1}{C-1})$, we achieve a rate reduction by a multiplicative factor
	\begin{equation*}
	\frac{R_{\text{uncoded}}^*(M)}{R(M)}= 1+\frac{r(C+1)(r-1)}{2(C-r-1)(C-1)}.
	\end{equation*}
	For a fixed $C$, this multiplicative factor increases as the access degree $r$ increases.\\
	b) The parameter $t=C-r$ corresponds to $\frac{M}{N} = \frac{1}{r}-\frac{1}{r\binom{C}{r}}$. We compare the rate in Theorem \ref{thm:scheme1} with a benchmark scheme obtained by combining the rate-memory trade-off of the MKR scheme and the memory-rate pair $(\frac{C}{r},0)$ (for the cyclic wrap-around MACC scheme, the achievability of $(\frac{C}{r},0)$ is shown in \cite{HKD}). Note that, for the combinatorial MACC scheme, the memory-rate pair $(\frac{C}{r},0)$ is achievable only with coded placement. For the ease of comparison, we assume that $C/r$ is an integer. Then, we have the rate achieved by the scheme in Theorem \ref{thm:scheme1}, $R(\frac{M}{N}) = \frac{1}{\binom{C}{r}}$ at $\frac{M}{N} = \frac{1}{r}-\frac{1}{r\binom{C}{r}}$. At the same $M/N$, the benchmark scheme achieves a rate $R_{\text{benchmark}}(\frac{M}{N}) =\frac{C}{r\binom{C}{r}} \frac{\binom{C}{\frac{C}{r}+r-1}}{\binom{C}{\frac{C}{r}-1}}$. Therefore, we have the following multiplicative reduction factor at $\frac{M}{N} = \frac{1}{r}-\frac{1}{r\binom{C}{r}}$,
	\begin{equation*}
		\frac{R_{\text{benchmark}}(\frac{M}{N})}{R(\frac{M}{N})} = \frac{C}{r} \frac{\binom{C}{\frac{C}{r}+r-1}}{\binom{C}{\frac{C}{r}-1}}
	\end{equation*}
	 where $\frac{R_{\text{benchmark}}(\frac{M}{N})}{R(\frac{M}{N})}>1$ since $\frac{C}{r}-1<\frac{C}{r}+r-1\leq C-(\frac{C}{r}-1)$. That is, in general $\frac{R_{\text{benchmark}}(\frac{M}{N})}{R(\frac{M}{N})}\approx C^{r-1}$ at the considered normalized cache memory value. 
\end{rem}

 Next, we present a combinatorial MACC scheme that performs better than the MKR scheme in the lower memory regime. In the following theorem, we present an achievability result for $ M\leq (N-\binom{C-1}{r})/C$. 

\begin{thm}
	\label{thm:scheme2}
	For the $(C,r,M,N)$ combinatorial MACC scheme with $N>\binom{C-1}{r}$, the rate $R(M) = N-CM$ is achievable for $0\leq M\leq (N-\binom{C-1}{r})/C$. 
\end{thm}
Proof of Theorem \ref{thm:scheme2} is given in Section \ref{proof:thm2}. \hfill$\blacksquare$

When the number of files $N$ is such that $\binom{C-1}{r}<N\leq \binom{C}{r}$, the rate $R(M) = N-CM$ is strictly less than the rate of the MKR scheme for $0\leq M\leq (N-\binom{C-1}{r})/C$.
\begin{rem}
	The scheme in Theorem \ref{thm:scheme2} achieves the rate $R(M)=\binom{C-1}{r}$ at $M=(\binom{C}{r}-\binom{C-1}{r})/C$, if $N=\binom{C}{r}$. At the same time the optimal rate under uncoded placement is $R^*_{\text{uncoded}}(M)=\binom{C-1}{r}(1+\frac{r}{C(r+1)})$. Therefore, when $N=\binom{C}{r}$, we have
	\begin{equation*}
		\frac{R^*_{\text{uncoded}}}{R(M)} = 1+\frac{r}{C(r+1)}
	\end{equation*}
at $(N-\binom{C-1}{r})/C$. That is, the rate reduction obtained from coded placement diminishes as the number of caches in the system increases. 
\end{rem}

Now, we present a lower bound on the optimal rate-memory trade-off for the $(C,r,M,N)$ combinatorial MACC scheme. 
\begin{thm}
	\label{lowerbound}
	For the $(C,r,M,N)$ combinatorial MACC scheme
	\begin{align}
	R^*(M) \geq \max_{\substack{s\in \{r,r+1,r+2,\hdots,C\} \\ \ell\in \left\{1,2,\hdots,\ceil{N/\binom{s}{r}}\right\}}} \frac{1}{\ell}&\Big\{N-\frac{\omega_{s,\ell}}{s+\omega_{s,\ell}}\Big(N-\ell\binom{s}{r}\Big)^+ \notag\\&- \Big(N- \ell\binom{C}{r}\Big)^+-sM\Big\} \label{lowereqn}
	\end{align}  
where $\omega_{s,\ell} = \min(C-s,\min\limits_i \binom{s+i}{r}\geq \ceil{\frac{N}{\ell}})$.
\end{thm}
Proof of Theorem \ref{lowerbound} is given in Section \ref{proof:thm3}. \hfill$\blacksquare$

The lower bound in Theorem \ref{lowerbound} has two parameters: $s$, which is related to the number of caches, and $\ell$, which is associated with the number of transmissions. To obtain this lower bound, we consider $s\in [r:C] $ caches and $\binom{s}{r}$ users who access caches only from those $s$ caches. From $\ceil{{N}/{\binom{s}{r}}}$ number of transmissions, all the $N$ files can be decoded at the $\binom{s}{r}$ users' end by appropriately choosing the demand vectors. Further, we split the total $\ceil{{N}/{\binom{s}{r}}}$ transmissions into $\ell$ transmissions and the remaining $\ceil{{N}/{\binom{s}{r}}}-\ell$ transmissions. Then, we bound the uncertainty in those two cases separately. This bounding technique is similar to the approach used in \cite{STC} and \cite{NaR2}, where lower bounds were derived for the dedicated cache scheme and the MACC scheme with cyclic wrap-around, respectively.

\begin{figure}[t]
	\captionsetup{justification = centering}
	\captionsetup{font=small,labelfont=small}
	\begin{center}
		\captionsetup{justification = centering}
		\includegraphics[width = 0.95\columnwidth]{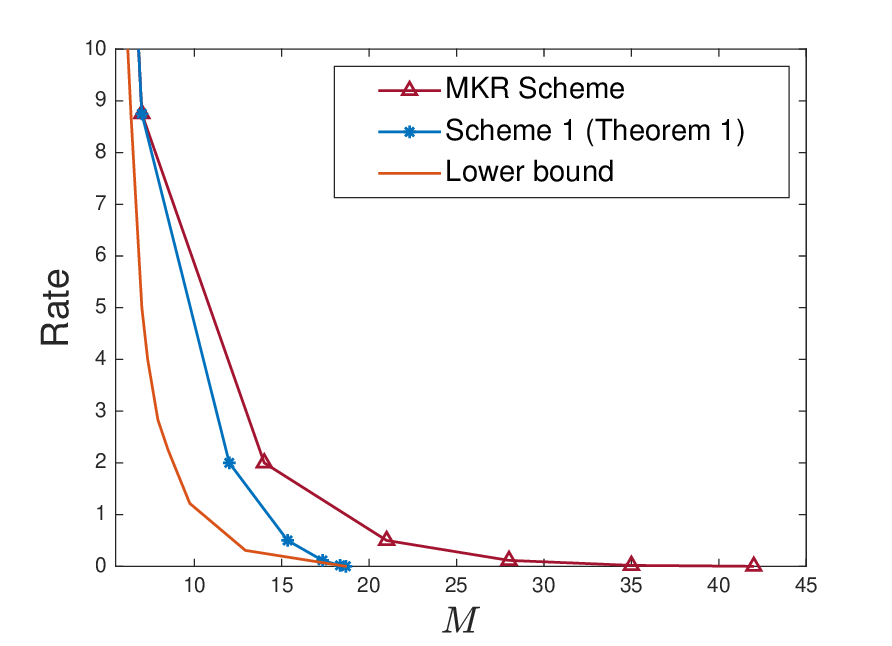}
		\caption{$(8,3,M,56)$ Combinatorial MACC scheme. A comparison between the MKR scheme and Scheme 1 ($M\geq 6$) }
		\label{plot1}
	\end{center}
\end{figure}

\begin{figure}[t]
	\captionsetup{justification = centering}
	\captionsetup{font=small,labelfont=small}
	\begin{center}
		\captionsetup{justification = centering}
		\includegraphics[width = 0.95\columnwidth]{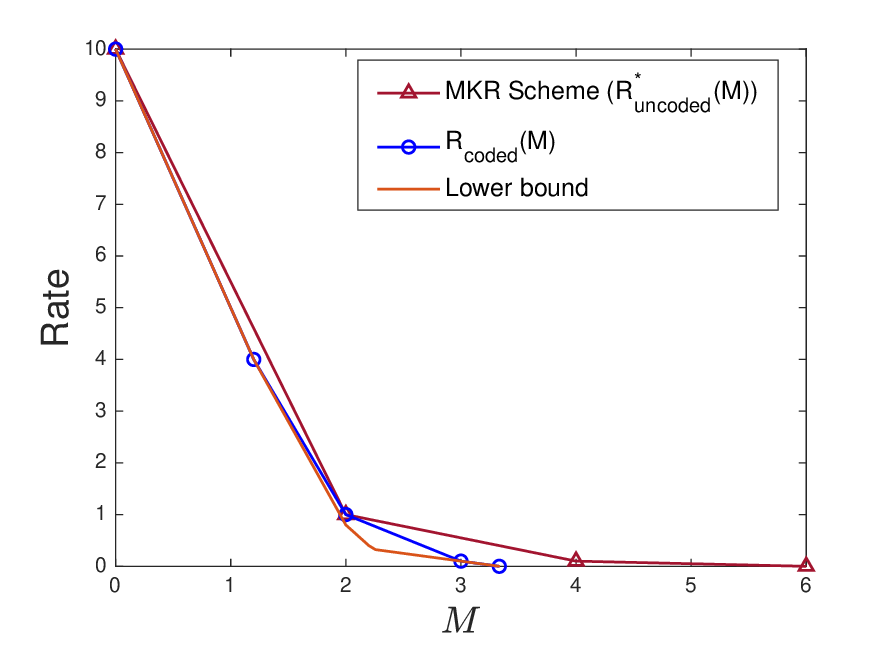}
		\caption{$(5,3,M,10)$ Combinatorial MACC scheme: A comparison between uncoded placement and coded placement }
		\label{plot3}
	\end{center}
\end{figure}

\begin{figure}[t]
	\captionsetup{justification = centering}
	\captionsetup{font=small,labelfont=small}
	\begin{center}
		\captionsetup{justification = centering}
		\includegraphics[width = 0.95\columnwidth]{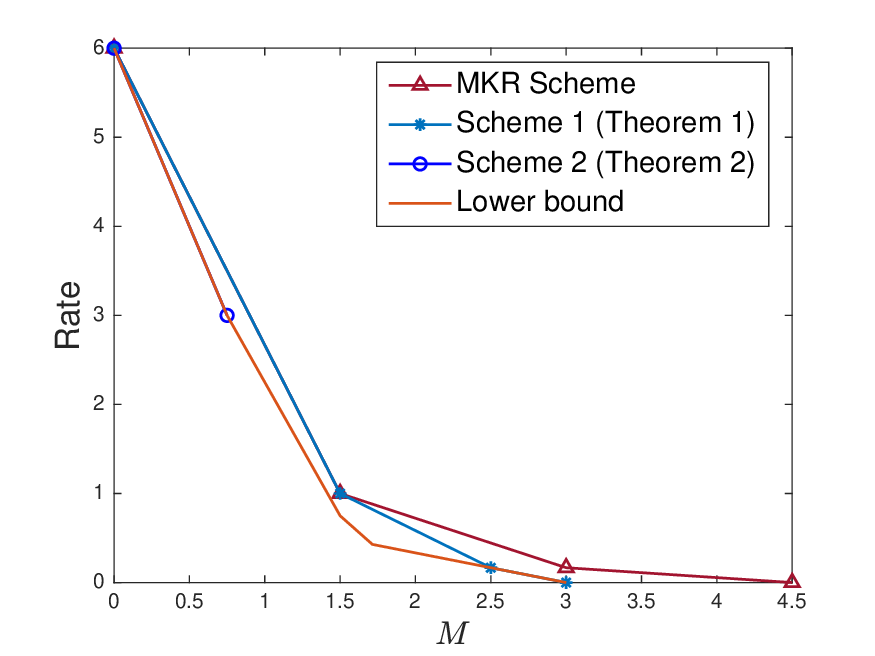}
		\caption{$(4,2,M,6)$ Combinatorial MACC scheme: The MKR scheme, Scheme 1, Scheme 2 and the lower bound }
		\label{plot2}
	\end{center}
\end{figure}

In Fig. \ref{plot1}, we plot the rate-memory trade-off in Theorem \ref{thm:scheme1} along with that of the MKR scheme. For $0\leq M\leq 7$ ($t=1$ gives $M=7$), both the schemes have the same rate-memory trade-off. However, for $M> 7$, our scheme has a strictly lower rate compared to the MKR scheme. It is worth noting that, in order to achieve $R(M)=0$, it is sufficient to have $M=N/r$, whereas, in the optimal scheme with uncoded placement, $M=N(C-r+1)/C$ is required to achieve $R(M)=0$. In Fig. \ref{plot3}, we compare the performance of the $(5,3,M,10)$ combinatorial MACC scheme under uncoded and coded placements. The rate-memory trade-off of the optimal scheme under uncoded placement (MKR scheme) is denoted as $R^*_{uncoded}(M)$, whereas $R_{coded}(M)$ is obtained from Theorem \ref{thm:scheme1} and Theorem \ref{thm:scheme2}, where the placement is coded. For the $(4,2,M,6)$ combinatorial MACC scheme, the improvement in rate using coded placement can be observed from Fig. \ref{plot2}. Also, notice that the rate-memory trade-off in Theorem \ref{thm:scheme1}, Theorem \ref{thm:scheme2} are optimal when $M\geq 2.5$, and $M\leq 0.75$, respectively.

Using the lower bound in Theorem \ref{lowerbound}, we show the following optimality results. First, we prove that the rate-memory trade-off in Theorem \ref{thm:scheme1} is optimal at a higher memory regime.
\begin{thm}
\label{optimality1}
For the $(C,r,M,N)$ combinatorial MACC scheme
\begin{equation}
R^*(M) = 1-\frac{rM}{N}
\end{equation}
for $\frac{\binom{C}{r}-1}{r\binom{C}{r}}\leq \frac{M}{N}\leq \frac{1}{r}$.
\end{thm}
Proof of Theorem \ref{optimality1} is given in Section \ref{proof:thm4}. \hfill$\blacksquare$

The optimality is shown towards the higher memory regime. When $\frac{\binom{C}{r}-1}{r\binom{C}{r}}\leq \frac{M}{N}\leq \frac{1}{r}$, the users would be requiring only a single subfile from the delivery phase and the server can meet that by a single coded transmission. Also, it is worth noting that in our proposed scheme (\textit{Scheme 1}), the delivery phase is optimal under the placement that we followed. Because the users in our proposed scheme and the users in the MKR scheme obtain the same subfiles from the respective placement phases. Thus, if we could design a better delivery phase, the same would apply to the MKR scheme. However, that is impossible since the MKR scheme is optimal for the placement employed. As the dedicated coded caching scheme (introduced in \cite{MaN}), the optimal rate-memory trade-off of the $(C,r,M,N)$ combinatorial MACC scheme is also open.

In the lower memory regime, when $M<N/C$, the rate-memory trade-off in Theorem \ref{thm:scheme1} is clearly not optimal, since the scheme in Theorem \ref{thm:scheme2} achieves a better rate. 
Now, we show that the rate-memory trade-off in Theorem \ref{thm:scheme2} is optimal if the number of users is not more than the number of files.
\begin{thm}
	\label{optimality2}
	For the $(C,r,M,N)$ combinatorial MACC scheme with $\binom{C-1}{r}<N\leq \binom{C}{r}$, we have
	\begin{equation}
	R^*(M) = N-CM
	\end{equation}
	for $0\leq M\leq \frac{N-\binom{C-1}{r}}{C}$.
\end{thm}
Proof of Theorem \ref{optimality2} is given in Section \ref{proof:thm5}. \hfill$\blacksquare$

In the proposed scheme, $N-\binom{C-1}{r}$ coded subfiles are kept in each cache (for $M=(N-\binom{C-1}{r})/C$) in the placement phase. However, the delivery phase is uncoded. That is, the server transmits uncoded subfiles in the delivery phase. A user receives $C-r$ subfiles (of the demanded file) directly, and the remaining $r$ subfiles need to be decoded from the $r$ accessible caches. A user requires $\binom{C-1}{r}$ subfiles from the delivery phase to decode a subfile of the demanded file from an accessible cache. That is, using $N-\binom{C-1}{r}$ coded subfiles in a cache and $\binom{C-1}{r}$ uncoded subfiles from the delivery phase, a user can decode $N$ subfiles, out of which only one is required for the user. Therefore, the ratio between the number of coded subfiles stored in a cache and the number of subfiles required to decode a subfile from a cache increases with $N$. Thus the optimality of the scheme is limited to the case where $N\leq \binom{C}{r}$.

Next, we show that the rate-memory trade-off of Scheme 1 is within a constant multiplicative factor from the optimal rate-memory trade-off, when $r=2$ and the number of files with the server is not less than the number of users in the system.
\begin{thm}
	\label{r=2gap}	
	For the $(C,r=2,M,N)$ combinatorial MACC scheme with $N\geq \binom{C}{2}$, we have
	\begin{equation}
	\label{gapeqn}
	\frac{R(M)}{R^*(M)}\leq 21
	\end{equation}
	where $R(M)$ is the achievable rate as defined in \eqref{rate}.
\end{thm}
Proof of Theorem \ref{r=2gap} is given in Section \ref{proof:thm6}. \hfill$\blacksquare$

In order to prove the optimality gap result, we divide the entire memory regime into three regions. For a lower memory regime (for $0\leq M\leq 2N/C$), we approximate the rate to $R(M=0) =  \binom{C}{2}$. In the second regime, where $2N/C\leq M\leq 0.1N$, we approximate the rate to $(N/M)^2$ (note that, $R(M)\leq (N/M)^2$). Finally, for $M\geq 0.1N$, we approximate the rate $R(M)\approx (N/M)^2(1-2M/N)$. To obtain the optimality gap result, we compute the ratio of these approximate rate-memory trade-off to the lower bound on $R^*(M)$ in Theorem \ref{lowerbound}. Due to these approximations and the analytical bounding techniques used, the optimality gap of 21 in Theorem \ref{r=2gap} is loose. From numerical simulations, for the $(C,r=2,M,N)$ combinatorial MACC scheme, we have
\begin{equation*}
\frac{R(M)}{R^*(M)}\leq 11.
\end{equation*}
 Also, the numerical simulations suggested that the multiplicative gap between the rate-memory trade-off given by Scheme 1 and the lower bound on $R^*(M)$ given in Theorem \ref{lowerbound} increases as $r$ increases. However, providing an analytical result on the optimality gap becomes increasingly hard with increasing $r$ because of the optimization problem in the lower bound expression \eqref{lowereqn}.
\begin{rem}
	To prove the optimality gap result for the combinatorial MACC scheme with $r=2$, we divide the entire parameter regime into 3 cases and further into several sub-cases. For a general $r$, due to the optimization problem in the lower bound \eqref{lowereqn} and the presence of two variables in the rate expression, namely $C$ and $r$, other than $M$ and $N$ make an analytical derivation of gap result increasingly hard. Also, numerical simulations suggest that the multiplicative factor (gap) between $R(M)$ given by Scheme 1 and the lower bound on $R^*(M)$ given in Theorem \ref{lowerbound} increases as $r$ increases. Additionally, for the $(C,r\leq C/2,M,N\geq \binom{C}{r})$ combinatorial MACC scheme with $uN/C\leq M\leq vN/r$, for some $u\geq 1$ and $v<1$, we have
	\begin{equation*}
		\frac{R(M)}{R^*(M)}\leq cz^rr!
	\end{equation*}
	where $c, z>1$ are constants (proof is given in Appendix \ref{AppendixY}). Similarly, for $M\geq vN/r$, we have shown in Appendix \ref{AppendixY} that 
	\begin{equation*}
		\frac{R(M)}{R^*(M)}\leq \left(\frac{r}{v}\right)^r.
	\end{equation*}
	Numerical simulations suggested that the multiplicative gap of 11 for $r=2$ increases to 15 and 55 as $r$ increases to 3 and 5, respectively. This indicates that the order provided by the analysis in Appendix \ref{AppendixY} is loose.  This is because of the fact that for a general $r$, we are unable to optimize the lower bound over the two variables, $s$ and $\ell$. However, the gap results indicate a possibility of the existence of either an achievable scheme with an additional gain that scales with $r$ or a better lower bound, which is higher than the lower bound in Theorem \ref{lowerbound}, approximately by a factor $r!$. In addition to that, if a better scheme exists, then that implies that a coded caching scheme with coded placement can perform better than the optimal scheme with uncoded placement by a multiplicative factor, which is not a constant and scales with $r$. 
\end{rem}

\section{Proofs}
\label{proofs}

\subsection{Proof of Theorem \ref{thm:scheme1}}
\label{proof:thm1}
In this section, we present a coding scheme (we refer to this scheme as \textit{Scheme 1}) that achieves the rate $\binom{C}{t+r}/\binom{C}{t}$ at cache memory $M$ (corresponding to every $t\in [C-r+1]$) as given in \eqref{thm1eqn}. First, let us consider the case $t\in [C-r]$ (for $t=C-r+1$, we present a separate coding scheme at the end of this proof). 

\noindent  
a) \textit{Placement phase}: The server divides each file into $\binom{C}{t}$ non-overlapping subfiles of equal size. Each subfile is indexed with a $t$-sized set $\mathcal{T}\subseteq [C]$. We assume $\mathcal{T}$ to be an ordered set. i.e., $\mathcal{T} = \{\tau_1,\tau_2,\dots,\tau_i,\dots,\tau_t\}$ with $\tau_1<\tau_2<\dots<\tau_t$. Also, whenever an ordering among the subfiles -indexed with $t$ sized subsets of $[C]$- is required, lexicographic ordering is followed. We have the file $W_n =\{W_{n,\mathcal{T}}: \mathcal{T}\subseteq [C],|\mathcal{T}|=t\} $ for every $n\in [N]$. Let us define $\tilde{r}\triangleq\min(r,t)$. Each subfile is further divided into $\tilde{r}!$ mini-subfiles. The subfile $W_{n,\mathcal{T}}$ is divided as follows:
\begin{equation*}
W_{n,\mathcal{T}}=\{W_{n,\mathcal{T}}^j: j\in [\tilde{r}!]\}.
\end{equation*}
Let $\mathbf{G}$ be a generator matrix of an $[\tilde{r}!t,\tilde{r}!]$ MDS code. For every $n\in [N]$ and for every $\mathcal{T}\subseteq [C]$ such that $|\mathcal{T}|=t$, the server encodes the mini-subfiles $(W_{n,\mathcal{T}}^j:j\in [\tilde{r}!])$ with $\mathbf{G}$ and obtains the coded mini-subfiles
\begin{equation}
\label{mdsround0}
(Y_{n,\mathcal{T}}^{1},Y_{n,\mathcal{T}}^{2},\dots,Y_{n,\mathcal{T}}^{\tilde{r}!t}) = (W_{n,\mathcal{T}}^j:j\in [\tilde{r}!])\mathbf{G}.
\end{equation}
Now, for an integer $c\in [C]$, and a set $\mathcal{T}\subseteq [C]$, $|\mathcal{T}|=t$ such that $\mathcal{T}\ni c$, we define a function $\phi_c^t:\{\mathcal{T}\subseteq[C]:\mathcal{T}\ni c,|\mathcal{T}|=t\}\rightarrow [t]$. The function $\phi_c^t(\mathcal{T})$ gives the position of $c$ in the ordered set $\mathcal{T}$ of size $t$. If $\mathcal{T} = \{\tau_1,\tau_2,\dots,\tau_i=c,\dots,\tau_t\}$, then  $\phi_c^t(\mathcal{T})=i$.

The placement phase consists of $\tilde{r}$ rounds- Round 0, Round 1, $\dots$, Round $\tilde{r}-1$. The content stored in cache $c$ in Round $b$ is denoted as $Z_c^b$, where $b\in \{0\} \cup [\tilde{r}-1]$. In Round 0, the server fills cache $c$ as
\begin{align*}
&Z_c^0 = \Big\{Y_{n,\mathcal{T}}^{q_\mathcal{T}^0+\ell_0}:\ell_0 \in [(\tilde{r}-1)!],\\&\hspace{2.5cm}\mathcal{T}\subseteq[C]:\mathcal{T}\ni c,|\mathcal{T}|=t,n\in [N]\Big\}
\end{align*} 
where $q_\mathcal{T}^0 = (\phi_c^t(\mathcal{T})-1)(\tilde{r}-1)!$. Notice that $(\tilde{r}-1)!\binom{C-1}{t-1}$ coded mini-subfiles (of all the files) are placed in all the caches in Round 0.

Now let us focus on the cache placement in Round $b$, where $b\in [\tilde{r}-1]$. In this round, the server further encodes certain coded mini-subfiles using an MDS code generator matrix. Let $\mathbf{G}^{(b)}$ be a systematic generator matrix of a $[2\binom{C-1}{t-1}-\sum_{i=1}^{b}\binom{r-1}{\tilde{r}-i}\binom{C-r}{t-\tilde{r}+i-1},\binom{C-1}{t-1}]$ MDS code, and let $\mathbf{G}^{(b)}=[\mathbf{I}_{\binom{C-1}{t-1}}|\mathbf{P}^{(b)}_{\binom{C-1}{t-1}\times \binom{C-1}{t-1}-\sum_{i=1}^{b}\binom{r-1}{\tilde{r}-i}\binom{C-r}{t-\tilde{r}+i-1}}]$. In Round $b$, the server picks the coded mini-subfiles $Y_{n,\mathcal{T}}^{q_\mathcal{T}^b+\ell_b}$ for every set  $\mathcal{T}\subseteq[C],|\mathcal{T}|=t$ such that $\mathcal{T}\ni c$, where $q_{\mathcal{T}}^b = \frac{\tilde{r}!}{\tilde{r}-b}t+\left(\phi_c^t(\mathcal{T})-1\right)\frac{\tilde{r}!}{(\tilde{r}-b)(\tilde{r}-b+1)}$ and $\ell_b\in \left[\frac{\tilde{r}!}{(\tilde{r}-b)(\tilde{r}-b+1)}\right]$. Those subfiles are encoded with $\mathbf{P}^{(b)}$ as follows:
\begin{align}
(Q_{n,c}^{\ell_b,1},Q_{n,c}^{\ell_b,2},\dots,Q_{n,c}^{\ell_b,\binom{C-1}{t-1}-\sum_{i=1}^{b}\binom{r-1}{\tilde{r}-i}\binom{C-r}{t-\tilde{r}+i-1}})& =\notag \\ \Big(Y_{n,\mathcal{T}}^{q_\mathcal{T}^b+\ell_b}: \mathcal{T}\subseteq[C],|\mathcal{T}|=t,\mathcal{T}&\ni c\Big)\mathbf{P}^{(b)}, \notag\\& \hspace{-4.7cm}\forall \ell_b\in \left[\frac{\tilde{r}!}{(\tilde{r}-b)(\tilde{r}-b+1)}\right], n\in [N]\label{mdsroundb}.
\end{align}	
We refer to the newly obtained coded mini-subfiles $Q_{n,c}^{\ell_b,j}$ as doubly-encoded mini-subfiles. In Round $b$, the server places the following doubly-encoded mini-subfiles in cache $c$,	  
\begin{align*}
Z_c^b =& \Big\{Q_{n,c}^{\ell_b,1},Q_{n,c}^{\ell_b,2},\dots,Q_{n,c}^{\ell_b,\binom{C-1}{t-1}-\sum_{i=1}^{b}\binom{r-1}{\tilde{r}-i}\binom{C-r}{t-\tilde{r}+i-1}}:\\&\hspace{2cm}\ell_b\in \left[\frac{\tilde{r}!}{(\tilde{r}-b)(\tilde{r}-b+1)}\right], n\in [N]\Big\}.
\end{align*} 
Note that, in Round $b$, a total of  $\frac{\tilde{r}!}{(\tilde{r}-b)(\tilde{r}-b+1)}\left(\binom{C-1}{t-1}-\sum_{i=1}^{b}\binom{r-1}{\tilde{r}-i}\binom{C-r}{t-\tilde{r}+i-1}\right)$ doubly-encoded mini-subfiles of all the files are kept in each cache.

The overall contents stored in cache $c$ in the placement phase is
\begin{equation*}
Z_c = \bigcup\limits_{b=0}^{\tilde{r}-1} Z_c^b, \hspace{0.2cm}\forall c\in [C].
\end{equation*}  
Each coded mini-subfile has $\frac{1}{\tilde{r}!\binom{C}{t}}$ of a file-size. Therefore, the normalized cache memory  is
\begin{align}
\frac{M}{N} &= \frac{(\tilde{r}-1)!\binom{C-1}{t-1})}{\tilde{r}!\binom{C}{t}}\notag\\&\hspace{0.5cm}+\frac{\sum\limits_{b=1}^{\tilde{r}-1}\frac{\tilde{r}!}{(\tilde{r}-b)(\tilde{r}-b+1)}\left(\binom{C-1}{t-1}-\sum\limits_{i=1}^{b}\binom{r-1}{\tilde{r}-i}\binom{C-r}{t-\tilde{r}+i-1}\right)}{\tilde{r}!\binom{C}{t}}\notag \\
&=\frac{t}{C}-\frac{1}{\binom{C}{t}}\sum\limits_{i=1}^{\tilde{r}-1}\frac{\tilde{r}-i}{r}\binom{r}{\tilde{r}-i+1}\binom{C-r}{t-\tilde{r}+i-1}\label{mbyn1}.
\end{align}
The calculation of $M/N$ value is elaborated in Appendix \ref{AppendixZ}.

\noindent \textit{b) Delivery phase:} Let $W_{d_\mathcal{U}}$ be the file demanded by user $\mathcal{U}$, where $\mathcal{U}\subseteq [C]$ and $|\mathcal{U}|=r$. During the delivery phase, the server makes a transmission corresponding to every $(t+r)$-sized subsets of $[C]$. The transmission corresponding to a set $\mathcal{S}\subseteq [C]$ such that $|\mathcal{S}|=t+r$ is 
\begin{equation*}
\bigoplus_{\substack{\mathcal{U}\subseteq \mathcal{S}\\|\mathcal{U}|=r}} W_{d_\mathcal{U},\mathcal{S}\backslash \mathcal{U}}.
\end{equation*} 
Note that, for a given $t\in [C]$, this delivery phase is the same as the delivery phase in the MKR scheme. There are $\binom{C}{t+r}$ number of $(t+r)$-sized subsets for $[C]$. The server makes transmission corresponding to each of those subsets. Therefore the number of coded subfiles transmitted in the delivery phase is $\binom{C}{t+r}$. Each subfile has $\frac{1}{\binom{C}{t}}$ of a file-size. Therefore the rate of transmission is ${\binom{C}{t+r}}/{\binom{C}{t}}$. 

Next, we show that by employing the above placement and delivery phases, all the users can decode their demanded files. Note that, all the MDS code generator matrices used for encoding the subfiles are known to the users as well. From the coded subfiles placed in the caches, if a user can decode all the subfiles that can be accessed by a corresponding user in the MKR scheme, then the decodability of the demanded files is guaranteed. This is because the delivery phase of our proposed scheme is the same as the delivery phase in the MKR scheme. Thus, we show that user $\mathcal{U}$ can decode the subfiles $W_{n,\mathcal{T}}$ such that $\mathcal{U}\cap \mathcal{T}\neq \phi$ from the coded subfiles stored in the accessible caches. That will ensure the decodability of the demanded files. First, we show that an arbitrary user $\mathcal{U}$ can decode $W_{n,\mathcal{T}}$ if $|\mathcal{T}\cap \mathcal{U}|=\tilde{r}$. Then we show that if the user decodes all the subfiles $W_{n,\mathcal{T}'}$ such that $|\mathcal{T}'\cap \mathcal{U}|>\tilde{r}-\beta$, then the user can decode the subfiles $W_{n,\mathcal{T}}$ with $|\mathcal{T}\cap \mathcal{U}|=\tilde{r}-\beta$, where $\beta \in [\tilde{r}-1]$. By sequentially applying $\beta=1$ to $\beta=\tilde{r}-1$, we have the required result that the user can decode all the subfiles $W_{n,\mathcal{T}}$ with $|\mathcal{T}\cap \mathcal{U}|\geq 1$.

\noindent \textit{c) Decodability:} Consider user $\mathcal{U}$ accessing cache $c$ for every $c\in \mathcal{U}$. Let us define $\tilde{r} \triangleq \min(r,t) $. Further, consider a $t$-sized set $\mathcal{T}\subseteq [C]$ such that $|\mathcal{T}\cap \mathcal{U}|=\tilde{r}$. The user has access to the following coded mini-subfiles of the subfile $W_{n,\mathcal{T}}$ from the caches: 
\begin{equation*}
\bigcup_{c\in \mathcal{T}\cap\mathcal{U}}\left\{\underbrace{Y_{n,\mathcal{T}}^{q_\mathcal{T}^0+\ell_0}}_{\text{available from }Z_c^0}:\ell_0\in [(\tilde{r}-1)!] \right\}.
\end{equation*} 
Note that the number of coded mini-subfiles of $W_{n,\mathcal{T}}$ accessible for the user is
\begin{equation*}
\mathrel{\bigg|} \bigcup_{c\in \mathcal{T}\cap\mathcal{U}}\left\{Y_{n,\mathcal{T}}^{q_\mathcal{T}^0+\ell_0}:\ell_0\in [(\tilde{r}-1)!] \right\}\mathrel{\bigg|}=(\tilde{r}-1)!\tilde{r}=\tilde{r}!.
\end{equation*}
Thus user $\mathcal{U}$ can decode $\tilde{r}!$ mini-subfiles of $W_{n,\mathcal{T}}$ from the $\tilde{r}!$ coded mini-subfiles (from \eqref{mdsround0}), and completely retrieve the subfile $W_{n,\mathcal{T}}$. That is, user $\mathcal{U}$ can decode every subfile $W_{n,\mathcal{T}}$ with $|\mathcal{T}\cap \mathcal{U}|=\tilde{r}$, using $Z_c^0$, $c\in \mathcal{U}$ alone. Also, note that the user gets all the coded mini-subfiles $\{Y_{n,\mathcal{T}}^{1},Y_{n,\mathcal{T}}^{2},\dots,Y_{n,\mathcal{T}}^{\tilde{r}!t}\}$ from $W_{n,\mathcal{T}}$ (from \eqref{mdsround0}), since $\mathbf{G}$ is known to all the users.

Now, we assume that user $\mathcal{U}$ has decoded all the subfiles $W_{n,\mathcal{T}'}$ with $|\mathcal{T}'\cap\mathcal{U}|> \tilde{r}-\beta$, where $\beta \in [\tilde{r}-1]$. Then, we show that the user can decode the subfiles $W_{n,\mathcal{T}}$ with $|\mathcal{T}\cap\mathcal{U}|=\tilde{r}-\beta$. From $Z_c^\beta$, user $\mathcal{U}$ has the doubly encoded mini-subfiles $\{Q_{n,c}^{\ell_\beta,1},Q_{n,c}^{\ell_\beta,2},\dots,Q_{n,c}^{\ell_\beta,\binom{C-1}{t-1}-\sum_{i=1}^{\beta}\binom{r-1}{\tilde{r}-i}\binom{C-r}{t-\tilde{r}+i-1}}\}$ for every $\ell_\beta\in [\frac{\tilde{r}!}{(\tilde{r}-\beta)(\tilde{r}-\beta+1)}]$. Since, the user decoded all the subfiles $W_{n,\mathcal{T}'}$ with $|\mathcal{T}'\cap\mathcal{U}|> \tilde{r}-\beta$, the user knows all the coded mini-subfiles in $\{Y_{n,\mathcal{T}'}^{q_{\mathcal{T}'}^\beta+\ell_\beta}: |\mathcal{T}'\cap\mathcal{U}|> \tilde{r}-\beta\}$. Note that, for a given $c\in \mathcal{U}$, we have
\begin{align*}
|\{\mathcal{T}': \mathcal{T}'\cap\mathcal{U}\ni c,|\mathcal{T}'\cap\mathcal{U}|> \tilde{r}-\beta\}|&=\\ \sum_{i=1}^{\beta}\binom{r-1}{\tilde{r}-i}&\binom{C-r}{t-\tilde{r}+i-1}.
\end{align*}
Therefore, using $\{Y_{n,\mathcal{T}'}^{q_{\mathcal{T}'}^\beta+\ell_\beta}: |\mathcal{T}'\cap\mathcal{U}|> \tilde{r}-\beta\}$ and $\{Q_{n,c}^{\ell_\beta,1},Q_{n,c}^{\ell_\beta,2},\dots,Q_{n,c}^{\ell_\beta,\binom{C-1}{t-1}-\sum_{i=1}^{\beta}\binom{r-1}{\tilde{r}-i}\binom{C-r}{t-\tilde{r}+i-1}}\}$, the user can decode the coded mini-subfiles $\{Y_{n,\mathcal{T}}^{q_\mathcal{T}^\beta+\ell_\beta}: |\mathcal{T}\cap\mathcal{U}|\ni c\}$ (from \eqref{mdsroundb}). Now, user $\mathcal{U}$ has the following coded mini-subfiles of the subfile $W_{n,\mathcal{T}}$:
\begin{align*}
&\bigcup\limits_{c\in \mathcal{T}\cap\mathcal{U}}\Big\{Y_{n,\mathcal{T}}^{q_\mathcal{T}^0+\ell_0},Y_{n,\mathcal{T}}^{q_\mathcal{T}^1+\ell_1},\dots,Y_{n,\mathcal{T}}^{q_\mathcal{T}^\beta+\ell_\beta}:\ell_0\in [(\tilde{r}-1)!],\\&\hspace{1.5cm} \ell_1\in [(\tilde{r}-2)! ],\dots,\ell_\beta\in [\frac{\tilde{r}!}{(\tilde{r}-\beta)(\tilde{r}-\beta+1)}]  \Big\}
\end{align*} 
where $|\mathcal{T}\cap\mathcal{U}|=\tilde{r}-\beta$.
Therefore, the number of coded mini-subfiles of $W_{n,\mathcal{T}}$ with the user is
\begin{align*}
&\bigg|\bigcup\limits_{c\in \mathcal{T}\cap\mathcal{U}}\Bigg\{Y_{n,\mathcal{T}}^{q_\mathcal{T}^0+\ell_0},Y_{n,\mathcal{T}}^{q_\mathcal{T}^1+\ell_1},\dots,Y_{n,\mathcal{T}}^{q_\mathcal{T}^\beta+\ell_\beta}:\ell_0\in [(\tilde{r}-1)!],\\&\hspace{1cm} \ell_1\in [(\tilde{r}-2)! ],\dots,\ell_\beta\in [\frac{\tilde{r}!}{(\tilde{r}-\beta)(\tilde{r}-\beta+1)}]  \Bigg\}\bigg| \\
&\hspace{1cm}=\left((\tilde{r}-1)!+\sum_{b=1}^\beta\frac{\tilde{r}!}{(\tilde{r}-b)(\tilde{r}-b+1)}\right)(\tilde{r}-\beta).
\end{align*}
The appropriate choice of $q_\mathcal{T}^i, i\in [\beta]$ makes sure that all the coded mini-subfiles are distinct. Now, we have
\begin{align*}
\sum_{b=1}^\beta\frac{\tilde{r}!}{(\tilde{r}-b)(\tilde{r}-b+1)}&=\tilde{r}!\left(\sum_{b=1}^\beta\frac{1}{\tilde{r}-b}-\frac{1}{\tilde{r}-b+1}\right)\\&=\tilde{r}!\left(\frac{1}{\tilde{r}-\beta}-\frac{1}{\tilde{r}}\right)\\&= (\tilde{r}-1)!\frac{\beta}{\tilde{r}-\beta}.
\end{align*}
Therefore, the user has 
\begin{align*}
&\bigg|\bigcup\limits_{c\in \mathcal{T}\cap\mathcal{U}}\bigg\{Y_{n,\mathcal{T}}^{q_\mathcal{T}^0+\ell_0},Y_{n,\mathcal{T}}^{q_\mathcal{T}^1+\ell_1},\dots,Y_{n,\mathcal{T}}^{q_\mathcal{T}^\beta+\ell_\beta}:\ell_0\in [(\tilde{r}-1)!],\\&\hspace{1cm} \ell_1\in [(\tilde{r}-2)! ],\dots,\ell_\beta\in [\frac{\tilde{r}!}{(\tilde{r}-\beta)(\tilde{r}-\beta+1)}]  \bigg\}\bigg| \\
&\hspace{1.5cm}=\left((\tilde{r}-1)!+(\tilde{r}-1)!\frac{\beta}{\tilde{r}-\beta}\right)(\tilde{r}-\beta)=\tilde{r}!
\end{align*}
coded mini-subfiles of $W_{n,\mathcal{T}}$. Thus, the user can retrieve the subfile $W_{n,\mathcal{T}}$. In other words, user $\mathcal{U}$ can decode all the subfiles $W_{n,\mathcal{T}}$ such that $|\mathcal{T}\cap\mathcal{U}|=\tilde{r}-\beta$. Since, this is true for every $\beta \in [\tilde{r}-1]$, the user gets all the subfiles $W_{n,\mathcal{T}}$ such that $|\mathcal{T}\cap\mathcal{U}|\neq \phi$ from the placement phase. That means, from the coded subfiles placed in the caches, a user can decode all the subfiles that can be accessed by a corresponding user in the MKR scheme. Thus, the decodability of the demanded files is guaranteed.

Finally, we consider the case $t=C-r+1$. The parameter $t=C-r+1$ corresponds to $M=N/r$ (shown in Appendix \ref{AppendixA}). In that case, it is possible to achieve rate $R(N/r)=0$ by placing the contents by properly employing coding. That is, from the accessible cache contents itself, all the users can decode the entire library of files. The cache placement is as follows. The server divides each file into $r$ non-overlapping subfiles of equal size. We have, $W_n = \{W_{n,1},W_{n,2},\dots,W_{n,r}\}$ for all $n\in [N]$. Let $\mathbf{G}_{r\times C}$ be a generator matrix of a $[C,r]$ MDS code. For every $n\in [N]$, the server does the following encoding procedure:
\begingroup
\allowdisplaybreaks
\begin{equation*}
(\tilde{W}_{n,1},\tilde{W}_{n,2},\dots,\tilde{W}_{n,C}) = (W_{n,1},W_{n,2},\dots,W_{n,r})\mathbf{G}.
\end{equation*}
Then the server fills the caches as 
\begin{equation*}
Z_c = \{\tilde{W}_{n,c}, \forall n\in [N]\}.
\end{equation*}
Therefore, user $\mathcal{U}$ gets access to $Z_c$ for every $c\in \mathcal{U}$. This means, the user has $r$ distinct coded subfiles $\tilde{W}_{n,c}$ for every $c\in \mathcal{U}$ and for all $n\in [N]$. From these coded subfiles, the user can decode all the files (from any $r$ coded subfiles, $r$ subfiles of a file can be decoded). This completes the proof of Theorem \ref{thm:scheme1}.\hfill$\blacksquare$
\endgroup

\begin{rem}
The proposed coding scheme depends upon $\min(r,t)$. The parameter $\tilde{r}=\min(r,t)$ is the maximum number of copies of a single subfile available -from the placement phase- for a user in the MKR scheme. When $t>r$, even though a subfile is stored in $t$ different caches, a user gets only a maximum of $r$ copies. In our scheme, we ensured that user $\mathcal{U}$ should be able to decode a subfile $W_{n,\mathcal{T}}$ using the accessible cache contents, if $1\leq |\mathcal{T}\cap\mathcal{U}|\leq \tilde{r}$. The number of rounds and the parameters of the encoding matrices depend upon this $\tilde{r}$. However, the general idea behind the coded placement remains the same irrespective of the value of $\tilde{r}$. 
\end{rem}

\subsection{Proof of Theorem \ref{thm:scheme2}}
\label{proof:thm2}
In this section, we present a coding scheme (we refer to this scheme as \textit{Scheme 2}).

First, we show the achievability of the rate $R(M) = \binom{C-1}{r}$ at $M = (N-\binom{C-1}{r})/C$ by presenting a coding scheme. 

\noindent \textit{a) Placement phase:} The server divides each file into $C$ non-overlapping subfiles of equal size. Thus, we have $W_n = \{W_{n,1},W_{n,2},\dots,W_{n,C}\}$ for all $n\in [N]$. Let $\mathbf{G}$ be a systematic generator matrix of a $[2N-\binom{C-1}{r},N]$ MDS code. Thus we can write $\mathbf{G}=[\mathbf{I}_N|\mathbf{P}_{N\times N-\binom{C-1}{r}}]$, where $\mathbf{I}_N$ is the identity matrix of size $N$. Then the server encodes the subfiles $(W_{1,i},W_{2,i},\dots,W_{N,i})$ using $\mathbf{P}_{N\times N-\binom{C-1}{r}}$, which results in $N-\binom{C-1}{r}$ coded subfiles. That is
\begin{align*}
&(Q_{1,i},Q_{2,i},\dots,Q_{N-\binom{C-1}{r},i}) =\\&\hspace{1.5cm} (W_{1,i},W_{2,i},\dots,W_{N,i})\mathbf{P}_{N\times N-\binom{C-1}{r}}, \hspace{0.2cm} \forall i\in [C].
\end{align*}
Then the server fills the caches as follows:
\begin{equation*}
Z_i = \{Q_{1,i},Q_{2,i},\dots,Q_{N-\binom{C-1}{r},i}\}, \hspace{.2cm} \forall i\in [C].
\end{equation*}
The $i^{\text{th}}$ cache contains $N-\binom{C-1}{r}$ linearly independent coded combinations of the subfiles $(W_{1,i},W_{2,i},\dots,W_{N,i})$. The size of a coded subfile is the same as the size of a subfile. Thus, we have the total size of a cache, $M = (N-\binom{C-1}{r})/C$.   

\noindent \textit{b) Delivery phase:} Let $W_{d_\mathcal{U}}$ be the file demanded by user $\mathcal{U}$ in the delivery phase. Let $n(\mathbf{d})$ denote the number of distinct files demanded by the users, i.e., $n(\mathbf{d}) = |\{W_{d_\mathcal{U}}:\mathcal{U}\subseteq [C],|\mathcal{U}|=r\}|$ If $n(\mathbf{d})\leq \binom{c-1}{r}$, then the server simply broadcasts those $n(\mathbf{d})$ files. Now, consider the case where $n(\mathbf{d})>\binom{C-1}{r}$. Let $n_i(\mathbf{d})$ denote the number of distinct files demanded by the users who do not access the $i^{\text{th}}$ cache, i.e.,
$n_i(\mathbf{d}) = |\{W_{d_\mathcal{U}}:\mathcal{U}\subseteq [C]\backslash\{i\},|\mathcal{U}|=r\}|$. Note that, $n_i(\mathbf{d})\leq \binom{C-1}{r}$ for all $i\in [C]$. If $n_i(\mathbf{d})=\binom{C-1}{r}$, then the server transmits $W_{d_\mathcal{U},i}$ for all $\mathcal{U}\subseteq [C]\backslash\{i\},$ such that $|\mathcal{U}|=r$. That is, the server transmits the $i^{\text{th}}$ subfile of the files demanded by the users who are not accessing cache $i$. If $n_i(\mathbf{d})<\binom{C-1}{r}$, then the server transmits the $i^{\text{th}}$ subfile of those distinct $n_i(\mathbf{d})$ files (files demanded by the users who do not access cache $i$) and the $i^{\text{th}}$ subfile of some other $\binom{C-1}{r}-n_i(\mathbf{d})$ files. The same procedure is done for all $i\in[C]$.    

Corresponding to every $i\in [C]$, the server transmits $\binom{C-1}{r}$ subfiles. Each subfile has $1/C$ of a file size. Thus the rate achieved is $\binom{C-1}{r}$. Now, the claim is that all the users can decode their demanded files completely. When $n(\mathbf{d})\leq \binom{C-1}{r}$, the decodability is trivial, as the demanded files are sent as such by the server. Let us consider the case where $n(\mathbf{d})> \binom{C-1}{r}$. Consider user $\mathcal{U}$ who accesses cache $i$ for all $i\in \mathcal{U}$. The user directly receives $W_{d_\mathcal{U},j}$ for all $j\in [C]\backslash \mathcal{U}$. Further, user $\mathcal{U}$ receives the $i^{\text{th}}$ subfile of some $\binom{C-1}{r}$ files, for every $i\in \mathcal{U}$. Also, for all $i\in \mathcal{U}$, the user has access to $N-\binom{C-1}{r}$ coded subfiles of the subfiles $\{W_{1,i},W_{2,i},\dots,W_{N,i}\}$  from cache $i$. Thus the user can decode $W_{n,i}$ for all $n\in [N]$ and $i\in \mathcal{U}$, since any $N$ columns of $\mathbf{G}$ are linearly independent. Therefore, user $\mathcal{U}$ gets $W_{d_\mathcal{U},k}$ for every $k\in [C]$. Hence, all the users can decode their demanded files. 

Further, if $M=0$, the rate $R=N$ is trivially achievable by simply broadcasting all the files. Thus, by memory sharing, the rate $R(M) = N-CM$ is achievable for $0\leq M\leq (N-\binom{C-1}{r})/C$. This completes the proof of Theorem \ref{thm:scheme2}. \hfill$\blacksquare$
\subsection{Proof of Theorem \ref{lowerbound}}
\label{proof:thm3}
The server has $N$ files $W_{[1:N]}$. Assume that the users are arranged in the lexicographic order of the subsets (of $[C]$ of size $r$) by which they are indexed. That is, $\{1,2,\dots,r\}$ will be the first in the order, $\{1,2,\dots,r-1,r+1\}$ will be the second and so on. Consider the set of first $s$ caches, where $s\in [r:C]$. The number of users who access caches only from this set is $\binom{s}{r}$. Let us focus on those $\binom{s}{r}$ users. Consider a demand vector $\mathbf{d}_1$ in which those $\binom{s}{r}$ users request for files $W_1,W_2,\dots,W_{\binom{s}{r}}$. That is, the first user (the user comes first in the lexicographic ordering among those $\binom{s}{r}$ users) demands $W_1$, the second user demands $W_2$, and so on. The remaining users request arbitrary files from $W_{[1:N]}$, which we are not interested in. Corresponding to $\mathbf{d}_1$, the server makes transmission $X_1$. From contents in the first $s$ caches $Z_{[1:s]}$ and transmission $X_1$, the considered $\binom{s}{r}$ users can collectively decode the first $\binom{s}{r}$ files (one file at each user). Now, consider a demand vector $\mathbf{d}_2$ in which those $\binom{s}{r}$ users request for files $W_{\binom{s}{r}+1},W_{\binom{s}{r}+2},\dots,W_{2\binom{s}{r}}$, and transmission $X_2$ corresponding to $\mathbf{d}_2$. From cache contents $Z_{[1:s]}$ and transmission $X_2$, those users can collectively decode the files $W_{[\binom{s}{r}+1:2\binom{s}{r}]}$. If we consider $\delta = \Big\lceil\frac{N}{\binom{s}{r}}\Big\rceil$ such demand vectors and the corresponding transmissions, all the $N$ files can be decoded at those $\binom{s}{r}$ user end. Thus, we have  
\begingroup
\allowdisplaybreaks
\begin{subequations}
	\begin{align}
	N& = H(W_{[1:N]})\leq H(Z_{[1:s]},X_{[1:\delta]})\\&=    H(Z_{[1:s]})+H(X_{[1:\delta]}|Z_{[1:s]})\label{lb1}.
	\end{align}
\end{subequations}
Consider an integer $\ell \in [1:\delta]$. We can expand \eqref{lb1} as,
\begin{subequations}
	\begin{align}
	N&\leq sM+H(X_{[1:\ell]}|Z_{[1:s]})+H(X_{[\ell+1:\delta]}|Z_{[1:s]},X_{[1:\ell]})\\
	&\leq sM+H(X_{[1:\ell]})+H(X_{[\ell+1:\delta]}|Z_{[1:s]},X_{[1:\ell]},W_{[1:\tilde{N}]})\label{Wls}
	\end{align}
\end{subequations}
\endgroup
where $\tilde{N} = \min(N,\ell \binom{s}{r})$. Using the cache contents $Z_{[1:s]}$ and transmissions $X_{[1:\ell]}$, the files $W_{[1:\tilde{N}]}$ can be decoded, hence \eqref{Wls} follows. Let us define, $\omega_{s,\ell} \triangleq \min(C-s,\min\limits_i \binom{s+i}{r}\geq \ceil{\frac{N}{\ell}})$. We can bound the entropy of $\ell$ transmissions by $\ell R^*(M)$, where each transmission rate is $R^*(M)$. Thus, we have
\begin{subequations}
	\begin{align}
	&N\leq sM+\ell R^*(M)+\notag\\&\qquad H(X_{[\ell+1:\delta]},Z_{[s+1:s+\omega_{s,\ell}]}|Z_{[1:s]},X_{[1:\ell]},W_{[1:\tilde{N}]})\label{lRstar}\\
	&\leq sM+\ell R^*(M)+\underbrace{H(Z_{[s+1:s+\omega_{s,\ell}]}|Z_{[1:s]},X_{[1:\ell]},W_{[1:\tilde{N}]})}_{\triangleq \mu}+\notag\\&\hspace{1.5cm}\underbrace{H(X_{[\ell+1:\delta]}|Z_{[1:s+\omega_{s,\ell}]},X_{[1:\ell]},W_{[1:\tilde{N}]})}_{\triangleq \psi} \label{mupsi}.
	\end{align}
\end{subequations}

Now, we find an upper bound on $\mu$ as follows:
\begin{subequations}
	\begin{align}
	\mu &= H(Z_{[s+1:s+\omega_{s,\ell}]}|Z_{[1:s]},X_{[1:\ell]},W_{[1:\tilde{N}]})\\
	&\leq H(Z_{[s+1:s+\omega_{s,\ell}]}|Z_{[1:s]},W_{[1:\tilde{N}]})\\&= H(Z_{[1:s+\omega_{s,\ell}]}|W_{[1:\tilde{N}]})-H(Z_{[1:s]}|W_{[1:\tilde{N}]}).\label{mueqn}
	\end{align}
\end{subequations}
By considering any $s$ caches from $[1:s+\omega_{s,\ell}]$, we can write an inequality similar to the one in \eqref{mueqn}.That is, 
\begin{equation}
\mu\leq H(Z_{[1:s+\omega_{s,\ell}]}|W_{[1:\tilde{N}]})-H(Z_{\mathcal{A}}|W_{[1:\tilde{N}]})\label{toavg}
\end{equation}
where $\mathcal{A}\subseteq [s+\omega_{s,\ell}],|\mathcal{A}|=s$, and $Z_{\mathcal{A}}$ is the contents stored in the caches indexed by the elements in the set $\mathcal{A}$. That means we can find $\binom{s+\omega_{s,\ell}}{s}$ such inequalities. Averaging over all those inequalities, we get
\begin{align}
\mu&\leq H(Z_{[1:s+\omega_{s,\ell}]}|W_{[1:\tilde{N}]})-\notag\\&\qquad\frac{1}{\binom{s+\omega_{s,\ell}}{s}}\sum_{\substack{\mathcal{A}\subseteq [s+\omega_{s,\ell}] \\ |\mathcal{A}|=s}} H(Z_{\mathcal{A}}|W_{[1:\tilde{N}]})\label{mubound}.
\end{align}
By applying  Lemma \ref{Entropy} for the random variables $Z_{[1:s+\omega_{s,\ell}]}$, we get
\begin{align*}
&\frac{1}{s+\omega_{s,\ell}} H(Z_{[1:s+\omega_{s,\ell}]}|W_{[1:\tilde{N}]})\leq\\&\qquad\qquad \frac{1}{\binom{s+\omega_{s,\ell}}{s}}\sum_{\substack{\mathcal{A}\subseteq [s+\omega_{s,\ell}] \\ |\mathcal{A}|=s}} \frac{H(Z_{\mathcal{A}}|W_{[1:\tilde{N}]})}{s}.
\end{align*}
Upon rearranging, we have
\begin{align}
\frac{1}{\binom{s+\omega_{s,\ell}}{s}}&\sum_{\substack{\mathcal{A}\subseteq [s+\omega_{s,\ell}] \\ |\mathcal{A}|=s}} H(Z_{\mathcal{A}}|W_{[1:\tilde{N}]})\geq\notag\\&\qquad\qquad \frac{s}{s+\omega_{s,\ell}} H(Z_{[1:s+\omega_{s,\ell}]}|W_{[1:\tilde{N}]})\label{mueqn2}.
\end{align}
Substituting \eqref{mueqn2} in \eqref{mubound}, we get
\begingroup
\allowdisplaybreaks
\begin{subequations}
	\begin{align*}
	\mu &\leq H(Z_{[1:s+\omega_{s,\ell}]}|W_{[1:\tilde{N}]})-\frac{s}{s+\omega_{s,\ell}} H(Z_{[1:s+\omega_{s,\ell}]}|W_{[1:\tilde{N}]})\\
	&= \frac{\omega_{s,\ell}}{s+\omega_{s,\ell}}H(Z_{[1:s+\omega_{s,\ell}]}|W_{[1:\tilde{N}]}).
	\end{align*}
\end{subequations}
\endgroup
Now, consider two cases a) if $\tilde{N} = \min(N,\ell \binom{s}{r})=N$, then
\begin{align}
H(Z_{[1:s+\omega_{s,\ell}]}|W_{[1:\tilde{N}]})=H(Z_{[1:s+\omega_{s,\ell}]}|W_{[1:N]})=0\label{mueqn3}
\end{align}
and b) if $\tilde{N} = \min(N,\ell \binom{s}{r})=\ell \binom{s}{r}$, then 
\begingroup
\allowdisplaybreaks
\begin{subequations}
	\begin{align}
	H(&Z_{[1:s+\omega_{s,\ell}]}|W_{[1:\tilde{N}]})=H(Z_{[1:s+\omega_{s,\ell}]}|W_{[1:\ell\binom{s}{r}]})\\&\leq H(Z_{[1:s+\omega_{s,\ell}]},W_{[\ell\binom{s}{r}+1:N]}|W_{[1:\ell\binom{s}{r}]})\\
	&= H(W_{[\ell\binom{s}{r}+1:N]}|W_{[1:\ell\binom{s}{r}]})+ H(Z_{[1:s+\omega_{s,\ell}]}|W_{[1:N]})\\
	&\leq H(W_{[\ell\binom{s}{r}+1:N]})\leq N-\ell\binom{s}{r}\label{mueqn4}
	\end{align}
\end{subequations}
\endgroup
where \eqref{mueqn3} and \eqref{mueqn4} follow from the fact that the cache contents are the functions of the files. Therefore, we have
\begin{align}
H(Z_{[1:s+\omega_{s,\ell}]}|W_{[1:\tilde{N}]})\leq \left(N-\ell \binom{s}{r}\right)^+.
\end{align}
Therefore, we have the following upper bound on $\mu$:
\begin{align}
\mu \leq \frac{\omega_{s,\ell}}{s+\omega_{s,\ell}}\left(N-\ell \binom{s}{r}\right)^+.\label{mu}
\end{align}
Now, we find an upper bound on $\psi$, where $\psi = H(X_{[\ell+1:\delta]}|Z_{[1:s+\omega_{s,\ell}]},X_{[1:\ell]},W_{[1:\tilde{N}]})$.	\\
We consider two cases a) if $N\leq \ell\binom{s+\omega_{s,\ell}}{r}$, then it is possible to decode all the files $W_{[1:N]}$ from $Z_{[1:s+\omega_{s,\ell}]}$ and $X_{[1:\ell]}$ by appropriately choosing the demand vectors $\mathbf{d}_1,\mathbf{d}_2,\dots,\mathbf{d}_\delta$. Then the uncertainty in $X_{[\ell+1:\delta]}$ is zero. That is, when $N\leq \ell \binom{s+\omega_{s,\ell}}{r}$, we have $\psi = 0$.\\
b) The second case $N> \ell\binom{s+\omega_{s,\ell}}{r}$ means that, $\omega_{s,\ell}=C-s$. Note that, $\omega_{s,\ell}$ is defined such that using the first $s+\omega_{s,\ell}$ caches and $\ell$ transmissions, it is possible to decode the remaining $N-\ell \binom{s}{r}$ files by appropriately choosing the demands of the remaining $\binom{s+\omega_{s,\ell}}{r}-\binom{s}{r}$ users, if $N\leq \ell\binom{C}{r}$. That is, $N> \ell\binom{s+\omega_{s,\ell}}{r}$ means that $N> \ell\binom{C}{r}$. Then, we have  
\begingroup
\allowdisplaybreaks
\begin{subequations}
	\begin{align}
	\psi &= H(X_{[\ell+1:\delta]}|Z_{[1:C]},X_{[1:\ell]},W_{[1:\tilde{N}]})\\
	&= H(X_{[\ell+1:\delta]}|Z_{[1:C]},X_{[1:\ell]},W_{[1:\ell\binom{C}{r}]})\label{psieqn1}\\
	&\leq H(X_{[\ell+1:\delta]},W_{[\ell\binom{C}{r}+1:N]}|Z_{[1:C]},X_{[1:\ell]},W_{[1:\ell\binom{C}{r}]})\\	
	&\leq H(W_{[\ell\binom{C}{r}+1:N]}|Z_{[1:C]},X_{[1:\ell]},W_{[1:\ell\binom{C}{r}]}) +\notag\\&\qquad\qquad\qquad H(X_{[\ell+1:\delta]}|Z_{[1:C]},X_{[1:\ell]},W_{[1:N]})\\
	&= H(W_{[\ell\binom{C}{r}+1:N]}|Z_{[1:C]},X_{[1:\ell]},W_{[1:\ell\binom{C}{r}]})\label{psieqn2}\\
	&\leq H(W_{[\ell\binom{C}{r}+1:N]})= N-\ell\binom{C}{r}.
	\end{align}
\end{subequations}
\endgroup
From cache contents $Z_{[1:C]}$, and the transmissions $X_{[1:\ell]}$ it is possible to decode the files $W_{[1:\ell\binom{C}{r}]}$, hence \eqref{psieqn1} follows. Further, \eqref{psieqn2} follows from the fact that given $W_{[1:N]}$, there is no uncertainty in the transmissions. Thus, we have the upper bound on $\psi$,
\begin{align}
\psi \leq \left(N-\ell\binom{C}{r}\right)^+ \label{psi}.
\end{align}
Substituting \eqref{mu} and \eqref{psi} in \eqref{mupsi}, we get
\begin{align*}
N \leq sM+\ell R^*(M)+X\frac{\omega_{s,\ell}}{s+\omega_{s,\ell}}&\left(N-\ell\binom{s}{r}\right)^+ +\\&\qquad \left(N-\ell\binom{C}{r}\right)^+.
\end{align*} 
Upon rearranging the terms, and optimizing over all the possible values of $s$ and $\ell$, we have the following lower bound on $R^*(M)$
\begin{align*}
R^*(M) &\geq \max_{\substack{s\in \{r,r+1,r+2,\hdots,C\} \\ \ell\in \left\{1,2,\hdots,\ceil{N/\binom{s}{r}}\right\}}} \frac{1}{\ell}\Big\{N-\\&\frac{\omega_{s,\ell}}{s+\omega_{s,\ell}}\left(N-\ell\binom{s}{r}\right)^+- \left(N-\ell\binom{C}{r}\right)^+-sM\Big\}
\end{align*}  
where $\omega_{s,\ell} = \min(C-s,\min\limits_i \binom{s+i}{r}\geq \ceil{\frac{N}{\ell}})$. This completes the proof of Theorem \ref{lowerbound}.\hfill $\blacksquare$

\subsection{Proof of Theorem \ref{optimality1}}
\label{proof:thm4}
First, we show that the rate $R(M) = 1-\frac{rM}{N}$ is achievable for $\frac{\binom{C}{r}-1}{r\binom{C}{r}}\leq \frac{M}{N}\leq \frac{1}{r}$. Substituting $t=C-r$ in \eqref{rate} gives $\frac{M}{N}=\frac{\binom{C}{r}-1}{r\binom{C}{r}}$, and $R = 1/\binom{C}{r}$. Similarly, $t=C-r+1$ gives $M/N=1/r$ and $R=0$. By memory sharing, the rate $R(M) = 1-\frac{rM}{N}$ is achievable for $\frac{\binom{C}{r}-1}{r\binom{C}{r}}\leq \frac{M}{N}\leq \frac{1}{r}$. Now, to show the converse, let us substitute $s=r$ and $\ell=N$ in \eqref{lowereqn}. That gives, 
\begin{equation*}
	R^*(M)\geq 1-\frac{rM}{N}.
\end{equation*}
Therefore, we can conclude that
\begin{equation*}
	R^*(M)= 1-\frac{rM}{N}
\end{equation*}
for $\frac{\binom{C}{r}-1}{r\binom{C}{r}}\leq \frac{M}{N}\leq \frac{1}{r}$. This completes the proof Theorem \ref{optimality1}. \hfill $\blacksquare$
\subsection{Proof of Theorem \ref{optimality2}}
\label{proof:thm5}
From Theorem \ref{thm:scheme2}, we know that for the $(C,r,M,N)$ combinatorial MACC scheme the rate 	
\begin{equation*}
	R(M) = N-CM
\end{equation*}
is achievable when $0\leq M\leq \frac{N-\binom{C-1}{r}}{C}$.
By substituting $s=C$ and $\ell=1$, we get
\begin{equation*}
	R^*(M) \geq N-CM.
\end{equation*}
Thus, we conclude that 
\begin{equation*}
	R^*(M) = N-CM
\end{equation*}
for $0\leq M\leq \frac{N-\binom{C-1}{r}}{C}$. \hfill$\blacksquare$
\subsection{Proof of Theorem \ref{r=2gap}}
\label{proof:thm6}
In this section, we show that the rate-memory trade-off of Scheme 1 is within a multiplicative gap of 21 from the lower bound on $R^*(M)$ in Theorem \ref{lowerbound}, when $r=2$ and $N\geq K$ (i.e., $N\geq \binom{C}{2}$). 
We consider the cases $C\leq 6$, $7\leq C\leq 11$, and $C\geq 12$ separately.\\
\textit{Case 1}: First assume that, $C\leq 6$. Then, we have
\begin{equation*}
	R(M)= \frac{\binom{C}{2}}{\binom{t+2}{2}}(1-\frac{2M}{N}).
\end{equation*}
By substituting $s=2$ and $\ell =N$ in \eqref{lowereqn}, we get
\begin{equation*}
	R^*(M)\geq 1-\frac{2M}{N}.
\end{equation*} 
Therefore, we obtain
\begin{equation}
	\frac{R(M)}{R^*(M)}\leq \frac{\binom{C}{2}}{\binom{t+2}{2}}\leq \binom{C}{2}\leq 15.\label{gap21-1}
\end{equation}

\noindent\textit{Case 2}: Now, we consider the case $7\leq C\leq 11$. For $M\geq N/C$ ($t=1$ corresponds to $M=N/C$), we have 
\begin{align*}
	R(M)&=\frac{\binom{C}{2}}{\binom{t+2}{2}}(1-\frac{2M}{N})\leq \frac{\binom{11}{2}}{\binom{1+2}{2}}(1-\frac{2M}{N})\\
	&\leq \frac{55}{3}(1-\frac{2M}{N}).
\end{align*}
Therefore, we have 
\begin{equation}
	\frac{R(M)}{R^*(M)}\leq \frac{55}{3}\leq 18.34 \label{gap21-2}
\end{equation}
for $M\geq N/C$, where $7\leq C\leq 11$.

\noindent Now, we consider the case $M\leq N/C$ and $7\leq C\leq 11$. By substituting $\ell=\ceil{\frac{N}{\binom{s}{2}}}$ in \eqref{lowereqn}, we get
\begin{equation}
	\label{gap1}
	R^*(M)\geq \frac{N-sM}{\ceil{\frac{N}{\binom{s}{2}}}}\geq \frac{N-sM}{\frac{N}{\binom{s}{2}}+1}\geq  \frac{\binom{s}{2}(1-\frac{sM}{N})}{1+\frac{\binom{s}{2}}{N}}.
\end{equation} 
Substituting $s=3$ in \eqref{gap1} yields
\begin{equation*}
	R^*(M)\geq \frac{3(1-\frac{3M}{N})}{1+\frac{3}{N}}.
\end{equation*} 
Since $C\geq 7$, we have $N\geq \binom{C}{2}\geq 21$. Therefore, we get 
\begin{equation*}
	R^*(M)\geq \frac{21}{8}(1-\frac{3M}{N}).
\end{equation*}
Therefore, we have 
\begin{equation*}
	\frac{R(M)}{R^*(M)}\leq \frac{55\times 8}{3\times 21}\frac{(1-\frac{2M}{N})}{(1-\frac{3M}{N})}.
\end{equation*}
However, we have $0\leq\frac{M}{N}\leq \frac{1}{C}\leq \frac{1}{7}$. Further, in that regime, the function $\frac{(1-\frac{2M}{N})}{(1-\frac{3M}{N})}$ is monotonically increasing in $\frac{M}{N}$, and we have
\begin{equation*}
	\frac{(1-\frac{2M}{N})}{(1-\frac{3M}{N})}\leq 1.25.
\end{equation*}
Therefore, we get
\begin{equation}
\label{gap21-3}
	\frac{R(M)}{R^*(M)}\leq \frac{55\times 8\times 1.25}{3\times 21}\leq 8.74.
\end{equation}
By combining \eqref{gap21-2} and \eqref{gap21-3}, we get
\begin{equation}
\label{gap21-11}
\frac{R(M)}{R^*(M)}\leq 18.34
\end{equation}
when $7\leq C\leq 11$.

\noindent \textit{Case 3}: Finally, we consider the third case, where $C\geq 12$. Now, we let $s = \floor{\mu C}$ and $\ell = \ceil{\nu N/\binom{s}{2}}$, where $\mu \in [2/C,1]$ and $\nu \in [0,1]$.
Then, we have 
\begin{align*}
R^*&(M) \geq \frac{1}{\ell}\bigg\{N-(1-\frac{s}{C})\left(N-\ell\binom{s}{2}\right)^+ -\\&\qquad\qquad\qquad\qquad\qquad \left(N-\ell\binom{C}{2}\right)^+-sM\bigg\}\\
&\geq \frac{N-(1-\frac{\floor{\mu C}}{C})\left(N-\ceil{\nu N/\binom{s}{2}}\binom{s}{2}\right)^+}{\ceil{\nu N/\binom{\floor{\mu C}}{2}}}-\\&\qquad\qquad\qquad\frac{\left(N-\ceil{\nu N/\binom{s}{2}}\binom{C}{2}\right)^+-\floor{\mu C}M}{\ceil{\nu N/\binom{\floor{\mu C}}{2}}}\\
& \geq \frac{N-(1-\frac{\mu C-1}{C})\left(N-\nu N\right)^+}{\nu N/\binom{\floor{\mu C}}{2}+1} -\\&\qquad\qquad\qquad \frac{\left(N-\ceil{\nu N/\binom{s}{2}}\binom{C}{2}\right)^+-\mu CM}{\nu N/\binom{\floor{\mu C}}{2}+1}\\
& \geq \frac{N-N(1-\mu+\frac{1}{C})(1-\nu )}{\nu N/\binom{\floor{\mu C}}{2}+1} -\\&\qquad\qquad\qquad \frac{N\left(1-\nu \binom{C}{2}/\binom{s}{2}\right)^+-\mu CM}{\nu N/\binom{\floor{\mu C}}{2}+1}\\
& \geq \frac{1-(1-\mu+\frac{1}{C})(1-\nu )}{\frac{2\nu }{\floor{\mu C}\floor{\mu C-1}}+\frac{1}{N}} -\\&\qquad\qquad\qquad \frac{\left(1-\nu \frac{C(C-1)}{\floor{\mu C}\floor{\mu C-1}}\right)^+-\mu C\frac{M}{N}}{\frac{2\nu }{\floor{\mu C}\floor{\mu C-1}}+\frac{1}{N}}\\
& \geq \frac{\nu+\mu(1-\nu)-\frac{1-\nu}{C} - \left(1-\frac{\nu}{\mu^2} \frac{(C-1)}{C-1/\mu}\right)^+-\mu C\frac{M}{N}}{\frac{2\nu }{(\mu C-1)(\mu C-2)}+\frac{1}{N}}.
\end{align*}   
We choose a value of $\nu\in [0,1]$ such that $\nu\geq \mu^2$. Then, we get $\left(1-\frac{\nu}{\mu^2} \frac{(C-1)}{C-1/\mu}\right)^+=0$. In that case, we have
\begin{align}
	R^*(M) & \geq \frac{\nu+\mu(1-\nu)-\frac{1-\nu}{C}-\mu C\frac{M}{N}}{\frac{2\nu }{(\mu C-1)(\mu C-2)}+\frac{1}{N}}\notag\\
	&\geq \frac{(C-\frac{1}{\mu})(C-\frac{2}{\mu})}{2}\frac{\nu+\mu(1-\nu)-\frac{1-\nu}{C}-\mu C\frac{M}{N}}{\frac{\nu }{\mu^2}+\frac{(C-\frac{1}{\mu})(C-\frac{2}{\mu})}{2N}}\label{rstar1}.
\end{align} 
Now, we assume that $0\leq M\leq 1.1N/C$. We have the rate $R(M)\leq \binom{C}{2}$.
Then, from \eqref{rstar1}, we obtain
\begin{align}
	\frac{R(M)}{R^*(M)}&\leq \frac{C(C-1)}{(C-\frac{1}{\mu})(C-\frac{2}{\mu})}\frac{\frac{\nu }{\mu^2}+\frac{(C-\frac{1}{\mu})(C-\frac{2}{\mu})}{2N}}{\nu+\mu(1-\nu)-\frac{1-\nu}{C}-1.1\mu}\notag\\
	&\leq \frac{1-\frac{1}{C}}{(1-\frac{1}{\mu C})(1-\frac{2}{\mu C})}\frac{\frac{\nu }{\mu^2}+\frac{(C-\frac{1}{\mu})(C-\frac{2}{\mu})}{2N}}{\nu+\mu(1-\nu)-\frac{1-\nu}{C}-1.1\mu}\label{gap2}.
\end{align}
Now, we substitute $\mu = 0.6$ and $\nu =1$ in \eqref{gap2}. Then, we obtain
\begin{align}
\label{gap21-4}
	\frac{R(M)}{R^*(M)}&\leq \frac{1}{(1-\frac{1}{12\times0.6 })(1-\frac{2}{12\times 0.6 })}\frac{\frac{1 }{.36}+1}{1-1.1\times 0.6}\leq 17.87
\end{align}
since $N\geq \binom{C}{2}\geq (C-1/\mu)(C-2/\mu)/2$. Next, we consider the case $1.1N/C\leq M\leq 1.9N/C$. Then, we have
\begin{align}
	R(M)&\leq R(1.1\frac{N}{C})=0.9 R(\frac{N}{C})+0.1R(\frac{2N}{C}) \label{memsharing}\\
	&\leq 0.9\frac{\binom{C}{3}}{C}+0.1\frac{\binom{C}{4}}{\binom{C}{2}}\notag\\&=0.9 \frac{(C-1)(C-2)}{6}+0.1\frac{(C-2)(C-3)}{12}\notag
\end{align}
where \eqref{memsharing} follows from memory sharing technique. Using \eqref{rstar1}, we get
\begin{align*}
	\frac{R(M)}{R^*(M)}&\leq \frac{2R(M)}{(C-\frac{1}{\mu})(C-\frac{2}{\mu})}\frac{\frac{\nu }{\mu^2}+\frac{(C-\frac{1}{\mu})(C-\frac{2}{\mu})}{2N}}{\nu+\mu(1-\nu)-\frac{1-\nu}{C}-1.9\mu}
\end{align*}
where
\begin{align*}
	&\frac{2R(M)}{(C-\frac{1}{\mu})(C-\frac{2}{\mu})}\leq \frac{0.9}{3} \frac{(C-1)(C-2)}{(C-\frac{1}{\mu})(C-\frac{2}{\mu})}+\\&\qquad\qquad\qquad\frac{0.1}{6}\frac{(C-2)(C-3)}{(C-\frac{1}{\mu})(C-\frac{2}{\mu})}\\
	&\qquad\leq \frac{0.9}{3} \frac{(1-\frac{1}{C})(1-\frac{2}{C})}{(1-\frac{1}{\mu C})(1-\frac{2}{\mu C})}+\frac{0.1}{6}\frac{(1-\frac{2}{C})(1-\frac{3}{C})}{(1-\frac{1}{\mu C})(1-\frac{2}{\mu C})}.
\end{align*}
Since $C\geq 12$, we have
\begin{align*}
	\frac{2R(M)}{(C-\frac{1}{\mu})(C-\frac{2}{\mu})}
	&\leq \frac{0.9}{3} \frac{1}{(1-\frac{1}{12\mu })(1-\frac{1}{6\mu })}+\\&\qquad\qquad\qquad\frac{0.1}{6}\frac{1}{(1-\frac{1}{12\mu })(1-\frac{1}{6\mu })}\\
	&=\frac{1.9}{6}\frac{1}{(1-\frac{1}{12\mu })(1-\frac{1}{6\mu })}.
\end{align*}
Now, let $\mu =0.37$ and $\nu=0.9969$. Then, we have
\begin{align}
	\frac{R(M)}{R^*(M)}&\leq\left(\frac{1.9}{6(1-\frac{1}{12\times 0.37 })(1-\frac{1}{6\times 0.37 })}\right)\times \\& \left(\frac{\frac{0.9969 }{0.37^2}+1}{0.9969+0.37\times 0.0031-\frac{0.0031}{12}-1.9\times 0.37}\right)\notag\\
	&\leq 20.895. \label{gap21-5}
\end{align}
Next, we consider the memory regime $1.9N/C\leq M\leq 0.0712N$. We have
\begin{align}
	R(M)&= \frac{\binom{C}{2}}{\binom{t+2}{2}}(1-\frac{2M}{N})\leq \frac{C(C-1)}{(t+2)(t+1)}\notag\\
	&\leq \frac{C(C-1)}{(\frac{CM}{N}+2)(\frac{CM}{N}+1)}\leq \left(\frac{N}{M}\right)^2.\label{approxrate}
\end{align}
We now let $s = \floor{0.6824N/M}$ and $\ell = \ceil{N/\binom{s}{2}}$. Then, we have the lower bound from \eqref{lowereqn}
\begin{align*}
R^*(M) &\geq \frac{1}{\ell}\bigg\{N-(1-\frac{s}{C})\left(N-\ell\binom{s}{2}\right)^+ -\\&\qquad\qquad\qquad\qquad\qquad \left(N-\ell\binom{C}{2}\right)^+-sM\bigg\}\\
&\geq \frac{N-(1-\frac{\floor{0.6824N/M}}{C})\left(N-\ceil{ N/\binom{s}{2}}\binom{s}{2}\right)^+}{\ceil{ N/\binom{\floor{0.6824N/M}}{2}}} -\\&\qquad\qquad \frac{\left(N-\ceil{ N/\binom{s}{2}}\binom{C}{2}\right)^+-\floor{0.6824N/M}M}{\ceil{ N/\binom{\floor{0.6824N/M}}{2}}}\\
&\geq \frac{N-0.6824N}{ \frac{N}{\binom{\floor{0.6824N/M}}{2}}+1}\\&\geq\frac{1-0.6824}{ \frac{2}{(0.6824N/M-1)(0.6824N/M-2)}+\frac{1}{N}}\\
&= \left(\frac{N}{M}\right)^2 \frac{0.3176}{ \frac{2}{(0.6824-M/N)(0.6824-2M/N)}+\frac{(\frac{N}{M})^2}{N}}.
\end{align*}  
Therefore, we obtain
\begin{align*}
	\frac{R(M)}{R^*(M)}&\leq\frac{ \frac{2}{(0.6824-M/N)(0.6824-2M/N)}+\frac{(\frac{N}{M})^2}{N}}{0.3176}.
\end{align*}
However, we have $N/M\leq C/1.9$ and $M/N\leq 0.0712$. Thus
\begin{align*}
	\frac{R(M)}{R^*(M)}&\leq\frac{ \frac{2}{(0.6824-M/N)(0.6824-2M/N)}+\frac{(\frac{N}{M})^2}{N}}{0.3176}\\
	&\leq\frac{ \frac{2}{(0.6824-.0712)(0.6824-.1424)}+\frac{C^2}{3.61N}}{0.3176}.
\end{align*}
Since $N\geq \binom{C}{2}$, we have
\begin{align*}
	\frac{C^2}{3.61N}&\leq \frac{C^2}{3.61\binom{C}{2}}=\frac{2}{3.61(1-\frac{1}{C})}\\&\leq \frac{2}{3.61(1-\frac{1}{12})}=0.6044.
\end{align*}
Therefore, we obtain
\begin{align}
\label{gap21-6}
	\frac{R(M)}{R^*(M)}
	&\leq\frac{ \frac{2}{(0.6824-.0712)(0.6824-.1424)}+0.6044}{0.3176}\leq 20.983.
\end{align}
Now, consider the case $M\geq 0.0712 N$. We have 
\begin{align*}
	R(M)&= \frac{\binom{C}{2}}{\binom{t+2}{2}}(1-\frac{2M}{N})\leq \frac{C(C-1)}{(t+2)(t+1)} \left(1-\frac{2M}{N}\right)\\
	&\leq \frac{C(C-1)}{(\frac{CM}{N}+2)(\frac{CM}{N}+1)} \left(1-\frac{2M}{N}\right)\\&\leq \left(\frac{N}{M}\right)^2 \left(1-\frac{2M}{N}\right).
\end{align*}
In \eqref{lowereqn}, substituting $\ell = \ceil{N/\binom{s}{2}}$ gives
\begin{align*}
	R^*(M) &\geq \frac{N-sM}{\ceil{\frac{N}{\binom{s}{2}}}}\geq \frac{N-sM}{\frac{N}{\binom{s}{2}}+1}= \frac{1-\frac{sM}{N}}{\frac{1}{\binom{s}{2}}+\frac{1}{N}}.
\end{align*} 
Therefore, we get
\begin{align*}
	\frac{R(M)}{R^*(M)}&\leq\left(\frac{1}{\binom{s}{2}}+\frac{1}{N}\right)\frac{(\frac{N}{M})^2(1-\frac{2M}{N})}{1-\frac{sM}{N}}.
\end{align*}
Also, we have $N\geq \binom{C}{2}\geq \binom{12}{2}=66$. Then, we obtain
\begin{equation}
	\label{eqstar2}
	\frac{R(M)}{R^*(M)}\leq\left(\frac{1}{\binom{s}{2}}+\frac{1}{66}\right)\frac{(\frac{N}{M})^2(1-\frac{2M}{N})}{1-\frac{sM}{N}}.
\end{equation}
Next consider the region $0.0712N\leq M\leq 0.1N$. By substituting $s =8$ in \eqref{eqstar2}, we get  
\begin{equation*}
	\frac{R(M)}{R^*(M)}\leq\left(\frac{1}{28}+\frac{1}{66}\right)\frac{(\frac{N}{M})^2(1-\frac{2M}{N})}{1-\frac{8M}{N}}
\end{equation*}
where $\frac{(\frac{N}{M})^2(1-\frac{2M}{N})}{1-\frac{8M}{N}}$ is a function of $\frac{M}{N}$ and has a maximum value 400 in the region $0.0712\leq M/N\leq 0.1$. Therefore, we have
\begin{equation}
	\frac{R(M)}{R^*(M)}\leq400\left(\frac{1}{28}+\frac{1}{66}\right)\leq 20.35.\label{gap21-7}
\end{equation}
Next consider the region $0.1N\leq M\leq 0.16N$. By substituting $s =5$ in \eqref{eqstar2}, we get  
\begin{equation*}
	\frac{R(M)}{R^*(M)}\leq\left(\frac{1}{10}+\frac{1}{66}\right)\frac{(\frac{N}{M})^2(1-\frac{2M}{N})}{1-\frac{5M}{N}}
\end{equation*}
where the function $\frac{(\frac{N}{M})^2(1-\frac{2M}{N})}{1-\frac{5M}{N}}$ has a maximum value 160 in the region $0.1\leq M/N\leq 0.16$. Therefore, we have
\begin{equation}
	\frac{R(M)}{R^*(M)}\leq160\left(\frac{1}{10}+\frac{1}{66}\right)\leq 18.43. \label{gap21-8}
\end{equation}
Next consider the region $0.16N\leq M\leq 0.25N$. By substituting $s =3$ in \eqref{eqstar2}, we get  
\begin{equation*}
	\frac{R(M)}{R^*(M)}\leq\left(\frac{1}{3}+\frac{1}{66}\right)\frac{(\frac{N}{M})^2(1-\frac{2M}{N})}{1-\frac{3M}{N}}
\end{equation*}
where the function $\frac{(\frac{N}{M})^2(1-\frac{2M}{N})}{1-\frac{3M}{N}}$ has a maximum value 51.082 in the region $0.16\leq M/N\leq 0.25$. Therefore, we have
\begin{equation}
	\frac{R(M)}{R^*(M)}\leq51.082\left(\frac{1}{3}+\frac{1}{66}\right)\leq 17.81. \label{gap21-9}
\end{equation}
Finally, consider the case $0.25N\leq M\leq 0.5N$. Substituting $s=2$ and $\ell=N$ in \eqref{lowereqn} yield
\begin{equation*}
	R^*(M)\geq 1-\frac{2M}{N}.
\end{equation*}
Also, we have
\begin{equation*}
	R(M)\leq \left(\frac{N}{M}\right)^2 \left(1-\frac{2M}{N}\right).
\end{equation*}
Therefore, we get 
\begin{equation}
	\frac{R(M)}{R^*(M)}\leq \left(\frac{N}{M}\right)^2\leq 16. \label{gap21-10}
\end{equation}
By combining \eqref{gap21-4},\eqref{gap21-5},\eqref{gap21-6},\eqref{gap21-7},\eqref{gap21-8},\eqref{gap21-9},and \eqref{gap21-10}, we get
\begin{equation}
\label{gap21-12}
\frac{R(M)}{R^*(M)}\leq 21
\end{equation}
when $C\geq 12$.

\noindent Finally, combining \eqref{gap21-1}, \eqref{gap21-11}, and \eqref{gap21-12} yields the required gap result
\begin{equation}
\label{gap21-13}
\frac{R(M)}{R^*(M)}\leq 21
\end{equation}
for the $(C,r=2,M,N\geq \binom{C}{2})$ combinatorial MACC scheme.
\hfill$\blacksquare$

\section{Conclusion}
\label{conclusion}
In this work, we presented two coding schemes for the combinatorial MACC setting introduced in \cite{MKR2}. Both the presented schemes employ coding in the placement phase in addition to the coded transmissions in the delivery phase. That is, we showed that with the help of a coded placement phase, it is possible to achieve a reduced rate compared to the optimal coded caching scheme under uncoded placement. Finally, we derived an information-theoretic lower bound on the optimal rate-memory trade-off of the combinatorial MACC scheme and showed that the first scheme is optimal at a higher memory regime and the second scheme is optimal when the number of files with the server is no more than the number of users in the system.


\begin{appendices}

		\section{}
	\label{AppendixZ}	
	In this section, we calculate the normalized cache memory as a continuation of \eqref{mbyn1}. From \eqref{mbyn1}, we have 	
	\begin{align*}
	\frac{M}{N} &= \frac{(\tilde{r}-1)!\binom{C-1}{t-1})}{\tilde{r}!\binom{C}{t}}+\\&\qquad\frac{\sum\limits_{b=1}^{\tilde{r}-1}\frac{\tilde{r}!}{(\tilde{r}-b)(\tilde{r}-b+1)}\left(\binom{C-1}{t-1}-\sum\limits_{i=1}^{b}\binom{r-1}{\tilde{r}-i}\binom{C-r}{t-\tilde{r}+i-1}\right)}{\tilde{r}!\binom{C}{t}}\\
	&=\frac{\binom{C-1}{t-1}\left((\tilde{r}-1)!+\sum\limits_{b=1}^{\tilde{r}-1}\frac{\tilde{r}!}{(\tilde{r}-b)(\tilde{r}-b+1)}\right)}{\tilde{r}!\binom{C}{t}}-\\&\qquad\qquad\frac{\sum\limits_{b=1}^{\tilde{r}-1}\frac{\tilde{r}!}{(\tilde{r}-b)(\tilde{r}-b+1)}\sum\limits_{i=1}^{b}\binom{r-1}{\tilde{r}-i}\binom{C-r}{t-\tilde{r}+i-1}}{\tilde{r}!\binom{C}{t}}.
	\end{align*}
	By expanding and cancelling the terms, we get 
	\begin{align*}
	\sum_{b=1}^{\tilde{r}-1}\frac{\tilde{r}!}{(\tilde{r}-b)(\tilde{r}-b+1)} &= \tilde{r}!\sum_{b=1}^{\tilde{r}-1}\left(\frac{1}{\tilde{r}-b}- \frac{1}{\tilde{r}-b+1}\right)\\
	& = \tilde{r}!\left(1- \frac{1}{\tilde{r}}\right)=(\tilde{r}-1)!(\tilde{r}-1).
	\end{align*}
	Thus, we have
	\begin{align*}
	\frac{M}{N}&=\frac{\tilde{r}!\binom{C-1}{t-1}-\sum\limits_{b=1}^{\tilde{r}-1}\frac{\tilde{r}!}{(\tilde{r}-b)(\tilde{r}-b+1)}\sum\limits_{i=1}^{b}\binom{r-1}{\tilde{r}-i}\binom{C-r}{t-\tilde{r}+i-1}}{\tilde{r}!\binom{C}{t}}.
	\end{align*}
	By changing the order of summation, we get 
	\begin{align*}
	\frac{M}{N}&=\frac{\tilde{r}!\binom{C-1}{t-1}-\sum\limits_{i=1}^{\tilde{r}-1}\binom{r-1}{\tilde{r}-i}\binom{C-r}{t-\tilde{r}+i-1}\sum\limits_{b=i}^{\tilde{r}-1}\frac{\tilde{r}!}{(\tilde{r}-b)(\tilde{r}-b+1)}}{\tilde{r}!\binom{C}{t}}.
	\end{align*}
	By expanding and cancelling the terms, we obtain 
	\begin{align*}
	\sum_{b=i}^{\tilde{r}-1}\frac{\tilde{r}!}{(\tilde{r}-b)(\tilde{r}-b+1)} &= \tilde{r}!\sum_{b=i}^{\tilde{r}-1}\left(\frac{1}{\tilde{r}-b}- \frac{1}{\tilde{r}-b+1}\right)\\
	&= \tilde{r}!\left(1- \frac{1}{\tilde{r}-i+1}\right)\\&=\tilde{r}!\frac{\tilde{r}-i}{\tilde{r}-i+1}.
	\end{align*}
	Therefore, we have the required expression in \eqref{thm1eqn} as follows:
	\begingroup
	\allowdisplaybreaks
	\begin{align*}
	\frac{M}{N}&=\frac{\tilde{r}!\binom{C-1}{t-1}-\tilde{r}!\sum\limits_{i=1}^{\tilde{r}-1}\frac{\tilde{r}-i}{\tilde{r}-i+1}\binom{r-1}{\tilde{r}-i}\binom{C-r}{t-\tilde{r}+i-1}}{\tilde{r}!\binom{C}{t}}\\
	&=\frac{t}{C}-\frac{1}{\binom{C}{t}}\sum\limits_{i=1}^{\tilde{r}-1}\frac{\tilde{r}-i}{r}\binom{r}{\tilde{r}-i+1}\binom{C-r}{t-\tilde{r}+i-1}.
	\end{align*}
	\endgroup \hfill $\blacksquare$
	

\section{}
\label{AppendixC}
In this section, we show that for an $M$ corresponds to a $t\in [C-r]$, as defined in \eqref{thm1eqn}, we can express the rate $R(M) = \binom{C}{r}\left(1-\frac{rM}{N}\right)/\binom{t+r}{r}$. From \eqref{thm1eqn}, we have

\begin{equation*}
\frac{M}{N}=
		\frac{t}{C}-\frac{1}{\binom{C}{t}}\sum\limits_{i=1}^{\tilde{r}-1}\frac{\tilde{r}-i}{r}\binom{r}{\tilde{r}-i+1}\binom{C-r}{t-\tilde{r}+i-1}.
\end{equation*}
Therefore, we get 
\begin{align*}
	&\frac{rM}{N}\binom{C}{t}=
	r\binom{C-1}{t-1}-\\&\qquad\qquad\qquad \sum\limits_{i=1}^{\tilde{r}-1}(\tilde{r}-i)\binom{r}{\tilde{r}-i+1}\binom{C-r}{t-\tilde{r}+i-1}.
\end{align*}
We expand the term 
\begin{align*}
	\sum\limits_{i=1}^{\tilde{r}-1}&(\tilde{r}-i)\binom{r}{\tilde{r}-i+1}\binom{C-r}{t-\tilde{r}+i-1}\\&=\sum\limits_{i=1}^{\tilde{r}-1}(\tilde{r}-i+1)\binom{r}{\tilde{r}-i+1}\binom{C-r}{t-\tilde{r}+i-1}-\\&\qquad\sum\limits_{i=1}^{\tilde{r}-1}\binom{r}{\tilde{r}-i+1}\binom{C-r}{t-\tilde{r}+i-1}\\
	&=\sum\limits_{j=2}^{\tilde{r}}j\binom{r}{j}\binom{C-r}{t-j}-\sum\limits_{j=2}^{\tilde{r}}\binom{r}{j}\binom{C-r}{t-j}
\end{align*}
where $j = \tilde{r}-i+1 $.

We can use the following identity to compute the above sum.  
\begin{lem}[\cite{Mes}]
	\label{lemmavander}
	Let $n_1,n_2$ be arbitrary positive integers. If $m$ is a positive integer such that $m\leq n_1+n_2$. Then
	\begin{equation*}
	\sum\limits_{k_1+k_2=m} k_1\binom{n_1}{k_1}\binom{n_2}{k_2}=\frac{mn_1}{n_1+n_2}\binom{n_1+n_2}{m}
	\end{equation*}
	where the summation ranges over all non-negative integers $k_1$ and $k_2$ such that $k_1\leq n_1$, $k_2\leq n_2$ and $k_1+k_2=m$.
\end{lem}
Lemma \ref{lemmavander} is a generalization of the Vandermonde identity:
\begin{equation}
\label{eqnvander}
\sum\limits_{k_1+k_2=m} \binom{n_1}{k_1}\binom{n_2}{k_2}=\binom{n_1+n_2}{m}.
\end{equation}
From Lemma \ref{lemmavander}, we have  
\begin{align*}
	\sum\limits_{j=2}^{\tilde{r}}j\binom{r}{j}\binom{C-r}{t-j} = \frac{rt}{C}\binom{C}{t}-r\binom{C-r}{t-1}.
\end{align*}
Similarly, from \eqref{eqnvander}, we get
\begin{align*}
	\sum\limits_{j=2}^{\tilde{r}}\binom{r}{j}\binom{C-r}{t-j} = \binom{C}{t}-r\binom{C-r}{t-1}-\binom{C-r}{t}.
\end{align*}
Therefore, we have 
\begin{align*}
	\frac{rM}{N}\binom{C}{t}&=
	r\binom{C-1}{t-1}-\bigg\{\frac{rt}{C}\binom{C}{t}-r\binom{C-r}{t-1}-\\&\qquad\left( \binom{C}{t}-r\binom{C-r}{t-1}-\binom{C-r}{t}\right)\bigg\}\\
	&=r\binom{C-1}{t-1}-\left(\frac{rt}{C}-1\right)\binom{C}{t}-\binom{C-r}{t}\\
	&=\binom{C}{t}-\binom{C-r}{t}.
\end{align*}
Thus, we get
\begin{equation}
	\frac{M}{N}=\frac{1}{r}\left(1-\frac{\binom{C-r}{t}}{\binom{C}{t}}\right)
	= \frac{1}{r}\left(1-\frac{\binom{C}{t+r}\binom{t+r}{t}}{\binom{C}{r}\binom{C}{t}}\right).\label{rnew}
\end{equation}
By rearranging \eqref{rnew}, we get
\begin{equation*}
	\frac{\binom{C}{t+r}}{\binom{C}{t}} = \frac{\binom{C}{r}\left(1-\frac{rM}{N}\right)}{\binom{t+r}{r}}.
\end{equation*}
We know that that $R(M) = \binom{C}{t+r}/\binom{C}{t}$. Therefore, we have the required rate expression
\begin{align*}
R(M) = \frac{\binom{C}{r}\left(1-\frac{rM}{N}\right)}{\binom{t+r}{r}}.
\end{align*}
\hfill $\blacksquare$
	\section{}
\label{AppendixA}
In this section, we show that, if we substitute $t=C-r+1$ in \eqref{thm1eqn}, we get $\frac{M}{N}=\frac{1}{r}$. We have
\begin{align}
\frac{M}{N}&=\frac{t}{C}-\frac{1}{r\binom{C}{\tilde{r}-1}}\sum\limits_{i=1}^{\tilde{r}-1}(\tilde{r}-i)\binom{r}{\tilde{r}-i+1}\binom{C-r}{t-\tilde{r}+r-i}.\notag
\end{align} 
a) Consider the case $\tilde{r}=r$. Then, we have
\begin{align}
\frac{M}{N}&=\frac{t}{C}-\frac{1}{r\binom{C}{r-1}}\sum\limits_{i=1}^{r-1}(r-i)\binom{r}{i-1}\binom{C-r}{r-i}\notag\\
&=\frac{C-r+1}{C}-\frac{1}{r\binom{C}{r-1}}\sum\limits_{i=1}^{r-1}(r-i)\binom{r}{i-1}\binom{C-r}{r-i}\notag.
\end{align}
From Lemma \ref{lemmavander}, we have 
\begin{equation*}
\sum\limits_{i=1}^{r-1}(r-i)\binom{r}{i-1}\binom{C-r}{r-i} = \frac{(r-1)(C-r)\binom{C}{r-1}}{C}.
\end{equation*}

Thus, we get 
\begin{align}
\frac{M}{N}&=\frac{C-r+1}{C}-\frac{(r-1)(C-r)\binom{C}{r-1}}{rC\binom{C}{r-1}} = \frac{1}{r}\notag.
\end{align}
b) Now, consider the case $\tilde{r}=t$. Then, we have
\begin{align}
\frac{M}{N}&=\frac{t}{C}-\frac{1}{r\binom{C}{t}}\sum\limits_{i=1}^{t-1}(t-i)\binom{r}{t-i+1}\binom{C-r}{i-1}\notag\\
&=\frac{t}{C}-\frac{\frac{rt}{C}\binom{C}{t}-\binom{C}{t}}{r\binom{C}{t}}\label{appeqn20}\\
&=\frac{1}{r}\notag
\end{align}
where \eqref{appeqn20} follows from Lemma \ref{lemmavander} and \eqref{eqnvander}. Thus we established that $t=C-r+1$ corresponds to $M/N=1/r$.\hfill $\blacksquare$

\section{}
\label{AppendixY}
In this section, we consider the rate-memory trade-off of the $(C,r,M,N\geq \binom{C}{r})$ combinatorial MACC scheme given by Scheme 1. We consider a general $C$ and an $r\leq C/2$, and consider the memory regime $u N/C\leq M\leq v N/r$, for some positive real numbers $u\geq 1$ and $v<1$. Notice that, for $r=2$, we chose $u=1.9$ and $v=0.1424$ (\textit{Case 2} in the proof of Theorem \ref{r=2gap}). For a $t\in [C-r+1]$ and corresponding $M$ from \eqref{thm1eqn}, we have
\begin{align}
R(M)&=\frac{\binom{C}{r}(1-\frac{rM}{N})}{\binom{t+r}{r}}\notag\\&\leq\frac{C(C-1)\dots(C-r+1)}{(\frac{CM}{N}+r)(\frac{CM}{N}+r-1)\dots(\frac{CM}{N}+1)}\notag\\&\leq \left(\frac{N}{M}\right)^r.\label{rateforr}
\end{align}
By substituting $\ell = \ceil{\frac{N}{\binom{s}{r}}}$ in \eqref{lowereqn}, we get
\begin{equation}
\label{lbr1}
R^*(M)\geq \frac{N-sM}{1+\frac{N}{\binom{s}{r}}} .
\end{equation}
Substituting $s = \floor{\alpha N/M}$ in \eqref{lbr1}, where $v<\alpha<1$ (ensures that $s\in [r:C]$), yields
 \begin{align}
 R^*(M)&\geq \frac{N-\floor{\alpha N/M}M}{1+\frac{N}{\binom{\floor{\alpha N/M}}{r}}}\geq \frac{N-\alpha N}{1+\frac{N}{\binom{\floor{\alpha N/M}}{r}}}\notag\\
 & \geq \frac{1-\alpha}{\frac{1}{N}+\frac{1}{\binom{\floor{\alpha N/M}}{r}}}\geq  \frac{1-\alpha}{\frac{1}{N}+\frac{r!}{(\frac{\alpha N}{M}-1)(\frac{\alpha N}{M}-2)\dots(\frac{\alpha N}{M}-r)}}\notag\\
 &\geq \left(\frac{N}{M}\right)^r  \frac{1-\alpha}{\frac{(\frac{N}{M})^r}{N}+\frac{r!}{(\alpha-\frac{ M}{N})(\alpha-\frac{ 2M}{N})\dots(\alpha-\frac{ rM}{N})}}   \notag\\
 &\geq \left(\frac{N}{M}\right)^r  \frac{1-\alpha}{\frac{(\frac{N}{M})^r}{\binom{C}{r}}+\frac{r!}{(\alpha-\frac{ M}{N})(\alpha-\frac{ 2M}{N})\dots(\alpha-\frac{ rM}{N})}}.         \label{lbr2}
 \end{align}
From \eqref{lbr1} and \eqref{lbr2}, we obtain
\begin{align}
\frac{R(M)}{R^*(M)}&\leq \frac{\frac{(\frac{N}{M})^r}{\binom{C}{r}}+\frac{r!}{(\alpha-\frac{ M}{N})(\alpha-\frac{ 2M}{N})\dots(\alpha-\frac{ rM}{N})}}{1-\alpha}\notag\\
&\leq \frac{\frac{(\frac{N}{M})^r}{\binom{C}{r}}+\frac{r!}{(\alpha-\frac{ rM}{N})^r}}{1-\alpha}.\label{lbr3}
\end{align}
Now, we separately bound the two terms in \eqref{lbr3} as follows:
\begin{align*}
\frac{(\frac{N}{M})^r}{\binom{C}{r}}\leq \frac{(C/u)^r}{\binom{C}{r}}= \frac{r!}{u^r(1-\frac{1}{C})(1-\frac{2}{C})\dots(1-\frac{r-1}{C})}.
\end{align*}
Since $r\leq C/2$, we have
\begin{align}
\label{lbr4}
\frac{(\frac{N}{M})^r}{\binom{C}{r}}\leq \frac{r!}{u^r(\frac{1}{2})^r}.
\end{align}
The second term in \eqref{lbr3} can be bounded as
\begin{align}
\frac{r!}{(\alpha-\frac{ rM}{N})^r}\leq\frac{r!}{(\alpha-v)^r}\label{lbr5}
\end{align}
since $M\leq vN/r$. By combining \eqref{lbr4} and \eqref{lbr5}, we get 
\begin{align}
\frac{R(M)}{R^*(M)}\leq \left(\frac{(\frac{2}{u})^r+\frac{1}{(\alpha-v)^r}}{1-\alpha}\right)r!.\label{lbr6}
\end{align}
From \eqref{lbr6}, we can approximately say that the gap grows in the order
\begin{align}
\frac{R(M)}{R^*(M)}\leq cz^rr!\label{lbr7} 
\end{align}
where $c$ and $z>1$ are constants. Also, note that by choosing $\alpha$ appropriately, we can bring $z$ close to 1 (though not arbitrarily close to 1). Thus, the optimality gap grows is by the (dominant) factor $r!$.

For $vN/r\leq M\leq N/r $, we have 
\begin{equation}
	\label{rateforr2}
	R(M)\leq \left(\frac{N}{M}\right)^r\left(1-\frac{rM}{N}\right).
\end{equation}
By substituting $s=r$ and $\ell = N$ in \eqref{lowereqn}, we get
\begin{equation}
	\label{lbr8}
	R^*(M)\geq 1-\frac{rM}{N} .
\end{equation}
Therefore, by combining \eqref{rateforr2} and \eqref{lbr8}, we obtain
\begin{equation}
	\label{gaprgen}
	\frac{R(M)}{R^*(M)}\leq \left(\frac{N}{M}\right)^r\leq \left(\frac{r}{v}\right)^r
\end{equation}
since $N/M\leq r/v$. In the considered memory regime, the gap grows with $r$ in the order of $(\frac{r}{v})^r$, where $v<1$.
\hfill $\blacksquare$
\end{appendices}

\begin{IEEEbiographynophoto}{K. K. Krishnan Namboodiri}
	(Graduate Student Member, IEEE) was born in Kerala, India. He received the B.Tech. degree in electronics and communication engineering from the National Institute of Technology, Calicut, in 2019. He is currently pursuing the Ph.D. degree with the Department of Electrical Communication Engineering, Indian Institute of Science, Bengaluru. His primary research interests include coded caching, index coding, information theory and wireless communication. He is a recipient of the Prime Minister's Research Fellowship (2021).
\end{IEEEbiographynophoto}

\begin{IEEEbiographynophoto}{B. Sundar Rajan}
	(Life Fellow, IEEE) was born in Tamil Nadu, India. He received the B.Sc. degree
	in mathematics from Madras University, Madras, India, in 1979, the B.Tech. degree in electronics from the Madras Institute of Technology, Madras, in 1982, and the M.Tech. and Ph.D. degrees in electrical engineering from IIT Kanpur, Kanpur, in 1984 and 1989, respectively.
	
	He was a Faculty Member at the Department of Electrical Engineering, IIT Delhi, New Delhi, from 1990 to 1997. He has been a Professor with the Department of Electrical Communication Engineering, Indian Institute of Science, Bengaluru, since 1998. His primary research interests include space-time coding for MIMO channels, distributed space-time coding and cooperative communication, coding for multiple-access and relay channels, and network coding.
	
	Dr. Rajan is a J. C. Bose National Fellow (2016–2025) and a member of the American Mathematical Society. He is a fellow of the Indian National Academy of Engineering, the Indian National Science Academy, the Indian Academy of Sciences, and the National Academy of Sciences, India. He was a recipient of Prof. Rustum Choksi Award by IISc for Excellence in Research in Engineering in 2009, the IETE Pune Center’s S. V. C. Aiya Award for Telecom Education in 2004, and the Best Academic Paper Award at IEEE WCNC in 2011. He served as the Technical Program Co-Chair for IEEE Information Theory Workshop (ITW’02) held in Bengaluru, in 2002. He was an Editor of IEEE \textsc{Wireless Communications Letters} (2012–2015) and IEEE \textsc{Transactions on Wireless Communications} (2007–2011), and an Associate Editor of Coding Theory for IEEE \textsc{Transactions on Information Theory} (2008–2011 and 2013–2015). 
\end{IEEEbiographynophoto}

\end{document}